\pdfoutput=1
 \documentclass[final,1p,times,twocolumn]{elsarticle}
 \usepackage{graphics}
\usepackage{amssymb}
\usepackage{url}
\usepackage{color}
\journal{Astroparticle Physics}

\begin{document}

\begin{frontmatter}



\title{Real-Time Supernova Neutrino Burst Monitor at Super-Kamiokande}


%\author{
%\input{authors-20130511}
%}
%\address{}

\author[icrr,ipmu]{K.~Abe}
\author[icrr]{Y.~Haga}
\author[icrr,ipmu]{Y.~Hayato}
\author[icrr,ipmu]{M.~Ikeda}
\author[icrr]{K.~Iyogi}
\author[icrr,ipmu]{J.~Kameda}
\author[icrr,ipmu]{Y.~Kishimoto}
\author[icrr,ipmu]{M.~Miura} 
\author[icrr,ipmu]{S.~Moriyama} 
\author[icrr,ipmu]{M.~Nakahata}
% ...
\address[icrr]{Kamioka Observatory, Institute for Cosmic Ray Research, University of Tokyo, Kamioka, Gifu 506-1205, Japan}
% ...
\address[ipmu]{Kavli Institute for the Physics and
Mathematics of the Universe (WPI), 
The University of Tokyo Institutes for Advanced Study,
University of Tokyo, Kashiwa, Chiba 277-8583, Japan}
% ...
\address{(The Super-Kamiokande Collaboration)}

\cortext[corr]{Corresponding Author, Hirokazu Ishino, Department of Physics, Okayama University, scishino@s.okayama-u.ac.jp}
\fntext[label-deceased]{Deceased.}

%\collaboration{The Super-Kamiokande Collaboration}
%(The Super-Kamiokande Collaboration)

\begin{abstract}
We present a real-time supernova neutrino burst monitor at Super-Kamiokande (SK).
Detecting supernova explosions by neutrinos in real time
is crucial for giving a clear picture of the explosion mechanism.
Since the neutrinos are expected to come earlier than light,
a fast broadcasting of the detection may give astronomers a chance to
make electromagnetic radiation observations of the explosions right at the onset.
The role of the monitor includes
a fast announcement of the neutrino burst detection to the world
and a determination of the supernova direction.
We present the online neutrino burst detection system and 
studies of the direction determination accuracy based on simulations at SK.
\end{abstract}


\end{frontmatter}

% \linenumbers

\section{Introduction}
\label{sec:introduction}

The detection of neutrinos from SN1987A opened a new era of
neutrino astronomy~\cite{sn1987a}.
Although the number of the detected neutrino events~\cite{footnote-on-event} was small,
significant information about the supernova (SN) explosion and
neutrino properties was obtained~\cite{sn1987a-derived}.
The current generation of detectors are waiting for the next SN neutrino
burst to accumulate event statistics much larger than those of SN1987A.

The SN burst neutrinos arrive at the Earth earlier than the electromagnetic radiation,
since the neutrinos generated at the core of the explosion
and emitted from the surface of the neutrinosphere 
%\textcolor{blue}{
travel at nearly the speed of light,%}
%seldom interact with the matter in the outer shells, 
while the shock waves propagating to the outside
with velocity much slower than the neutrino velocity result in the emission of electromagnetic radiation~\cite{sm-adam}.
%The delay may depend on the structure of the envelope of the core, and
%is expected to last from a few hours to days.
The delay may depend on the structure of the envelope of the core as well
as the surrounding stellar environment, and is expected to range between
tens of minutes and tens of hours~\cite{sn-outburst}.
Therefore, the detection of the neutrino burst can generate a warning
able to allow the astronomers to observe the radiation from the onset of the
explosion.
Such warning systems have been developed by 
several neutrino observatories~\cite{kam-imb}~\cite{sn-burst-monitor}
as well as by the supernova early warning system (SNEWS)~\cite{snews}.

It is also important to determine the SN direction using the neutrino signal:
the direction information can guide optical instruments toward
the SN explosion and enable observation of the onset of radiation.
Among the neutrino detectors operating at present, Super-Kamiokande (SK)
is the only detector able to determine the SN direction using neutrino events.
We have developed a SN direction determination method by applying a maximum likelihood fit.

SK is the world's largest water Cherenkov detector located 1,000~m underground, inside
a mountain in Kamioka, Gifu, Japan.
The detector consists of 50,000 tons of ultra-pure water 
and about 13,000 photomultiplier tubes (PMTs).
Based on the information of the yields and the arrival timing of Cherenkov photons
for individual PMTs as well as the PMT locations,
SK is able to measure the position, direction and energy of a neutrino event in real time.
Details of the SK detector and its performance are described in~\cite{sk-detector}.
In 2008, SK upgraded its readout system.
The system has improved the data processing speed significantly, lowering the trigger
energy threshold and dead time for the SN burst events~\cite{qbee}.

We will describe the SN neutrino burst monitor at SK.
In Section~\ref{sec:sn-monitor}, we will describe details of the monitor system
and its performance.
We first describe SN models used in this report for performance evaluations.
A Monte-Carlo (MC) simulation with the SN models is utilized for the estimation 
of the detection efficiency.
We will explain the selection criteria to discriminate SN bursts
and the main background in SK: radioactive decays caused by
cosmic ray muon spallation~\cite{spallation}.  
In Section~\ref{sec:direction}, we will describe a method to reconstruct
the SN direction and studies of its performance.

\section{Real time supernova neutrino burst monitor}
\label{sec:sn-monitor}

\subsection{Supernova models}
\label{sec:SN-MC}

We first describe the SN models we use in this report.
We employ two models for the SN neutrino burst:
the Wilson model~\cite{wilson-model}
and the Nakazato model~\cite{nakazato-model}.
For the Nakazato model, we choose two parameter sets:
$M=20$, $t_{\rm revive}=200$~msec and $Z=0.02$ (NK1), 
and $M=13$, $t_{\rm revive}=100$~msec and $Z=0.004$ (NK2),
where $M$ is the progenitor mass in units of the solar mass,
$t_{\rm revive}$ is the shock revival time and $Z$ is the metallicity,
respectively.
We choose the first one for the SN1987A progenitor mass, which is about 20 times 
the solar mass.
The model with the latter parameters gives the smallest neutrino fluxes 
in the Nakazato model.
Both models provide time dependences of neutrino luminosities and energy spectra for
$\nu_e$, $\bar{\nu}_e$ and $\nu_x$,
for 18 seconds (20 seconds) for the Wilson (NK) model,  
where $\nu_x$ refers to the muon and tau types of neutrinos and anti-neutrinos.
The anti-electron neutrino fluences of the models of Wilson, NK1 and NK2
are $16.0$, $9.8$ and $9.4$, respectively, in units of $10^{10}/$cm$^2$
in the energy range of 7 to 50~MeV at the distance of 10~kpc without neutrino
oscillation.
We also take into account neutrino oscillations based on~\cite{sn-nu-osci}.
We assume $P_H=0$ in the parameterization of \cite{sn-nu-osci}, which implies
adiabatic transitions between electron and tau (anti-) neutrinos 
due to $\sin^2 2\theta_{13}=0.095\pm 0.010$~\cite{pdg}.
In this report, we do not take into account collective effects.
We use the cross sections~\cite{ibd-crx} for the inverse beta decays,
\cite{cc-oxygen}\cite{cc-oxygen2} for the charged current interactions to oxygen,
and \cite{elastic} for the electron elastic scatterings.

%Table ~\ref{table:expected-events} shows the expected numbers of events at SK
%in the 22.5~kton fiducial volume with the total energy threshold of 7~MeV,
%which is the event selection condition for the SN burst monitor,
%for a SN with a distance of 10~kpc.
%Figure~\ref{fig:sk_vis_ene} shows the reconstructed energy distributiosn at SK
%for the Wilson and NK1 models without neutrino oscillations.

Monte-Carlo simulation samples are generated for the three SN models and
three neutrino oscillation hypotheses
by making use of the full SK detector MC simulator based on Geant3
%\textcolor{blue}{
in the SK inner detector volume (32.5-kton) in the energy range
of 3 to 60~MeV taking into account the trigger threshold curve.%}
The calibration of the detector and its simulation are described in~\cite{sk-calibration}.
We apply the SK standard reconstruction program to the generated MC events
to obtain the vertex position, the direction and the total energy of each event.
In the SN monitor, we use events with total energy greater than 7~MeV
in the 22.5-kton fiducial volume, 
where the fiducial volume is defined as the volume
whose surface is located 2~m inside from the surface of the SK inner detector volume.

\begin{figure}[h]
\begin{center}
\includegraphics[width=65mm,clip]{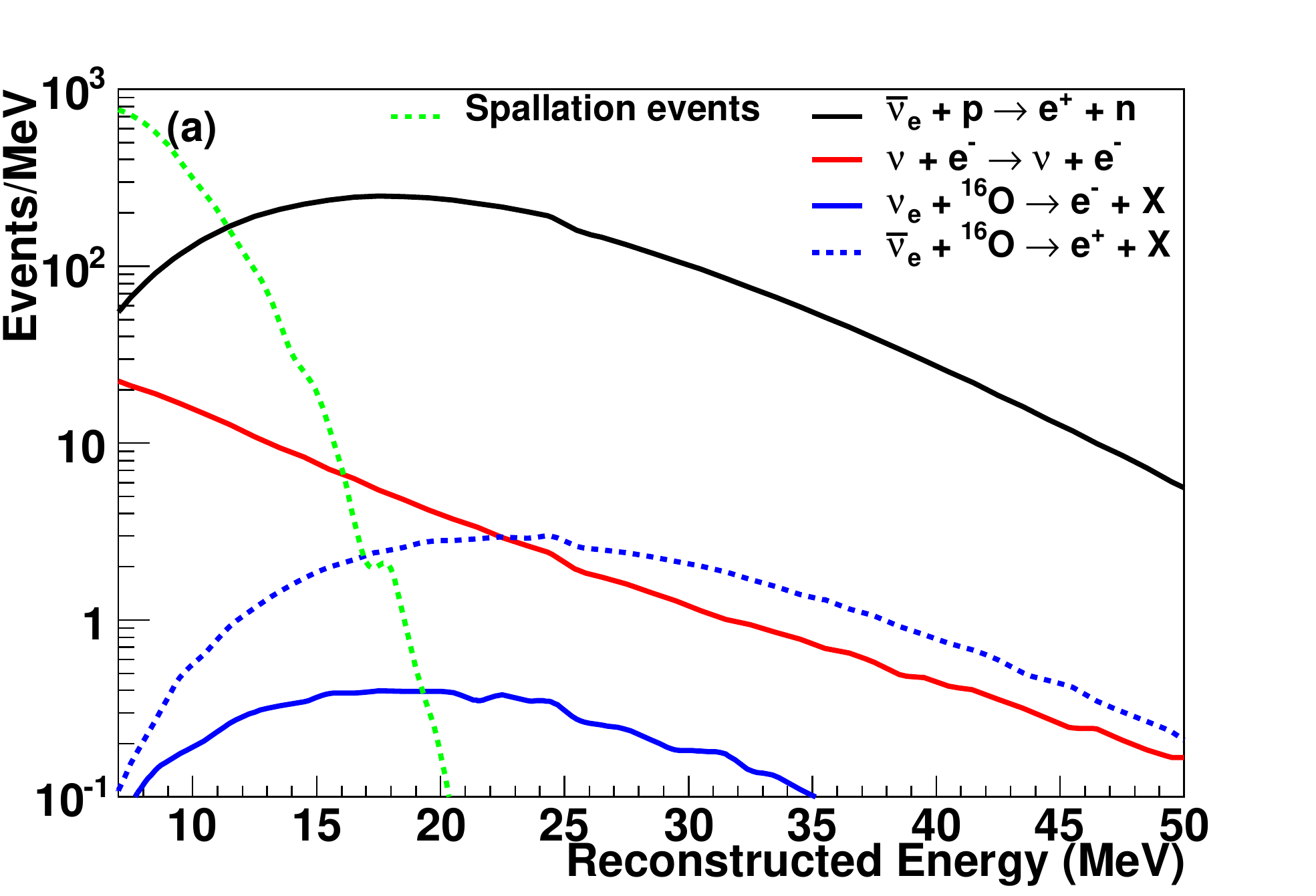}
\includegraphics[width=65mm,clip]{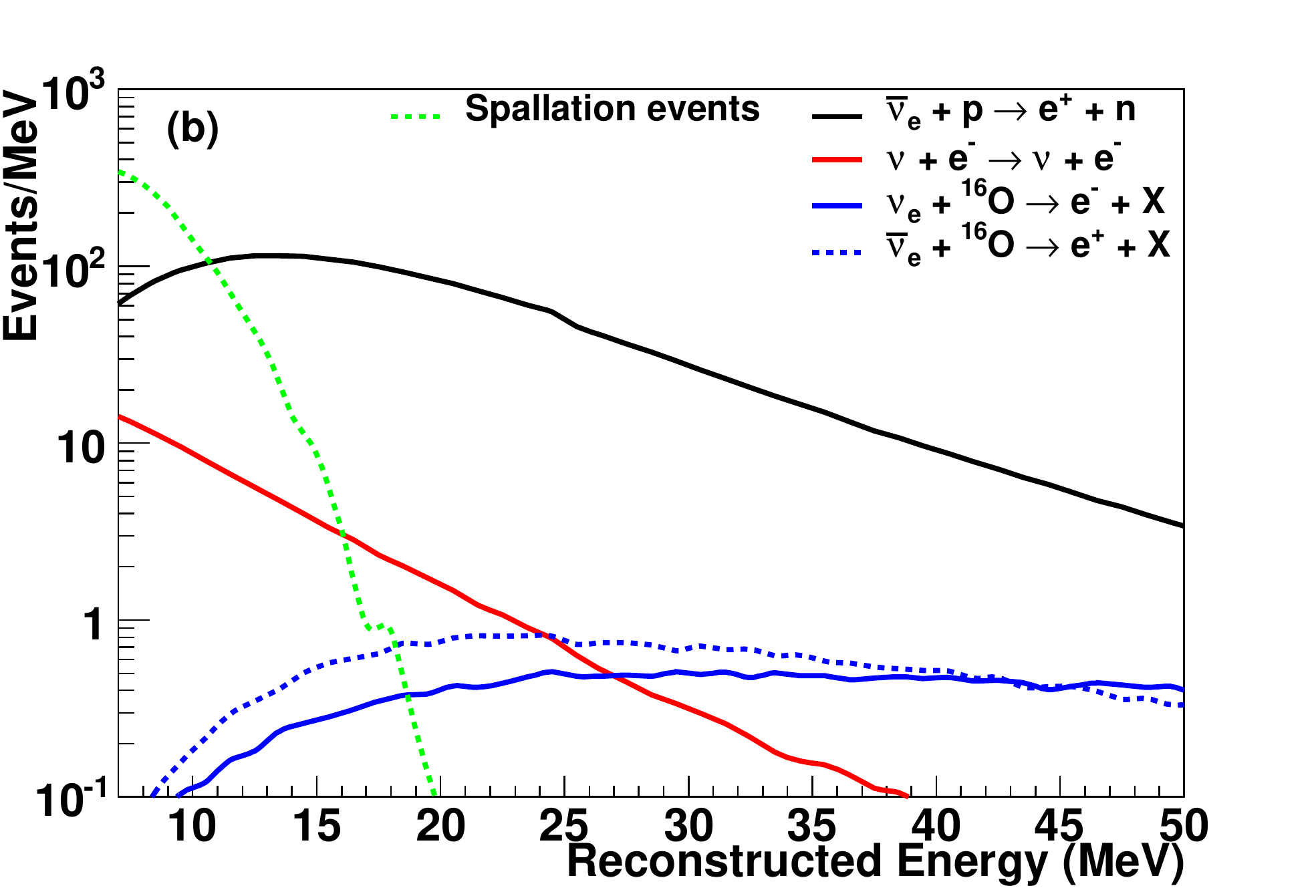}
\includegraphics[width=65mm,clip]{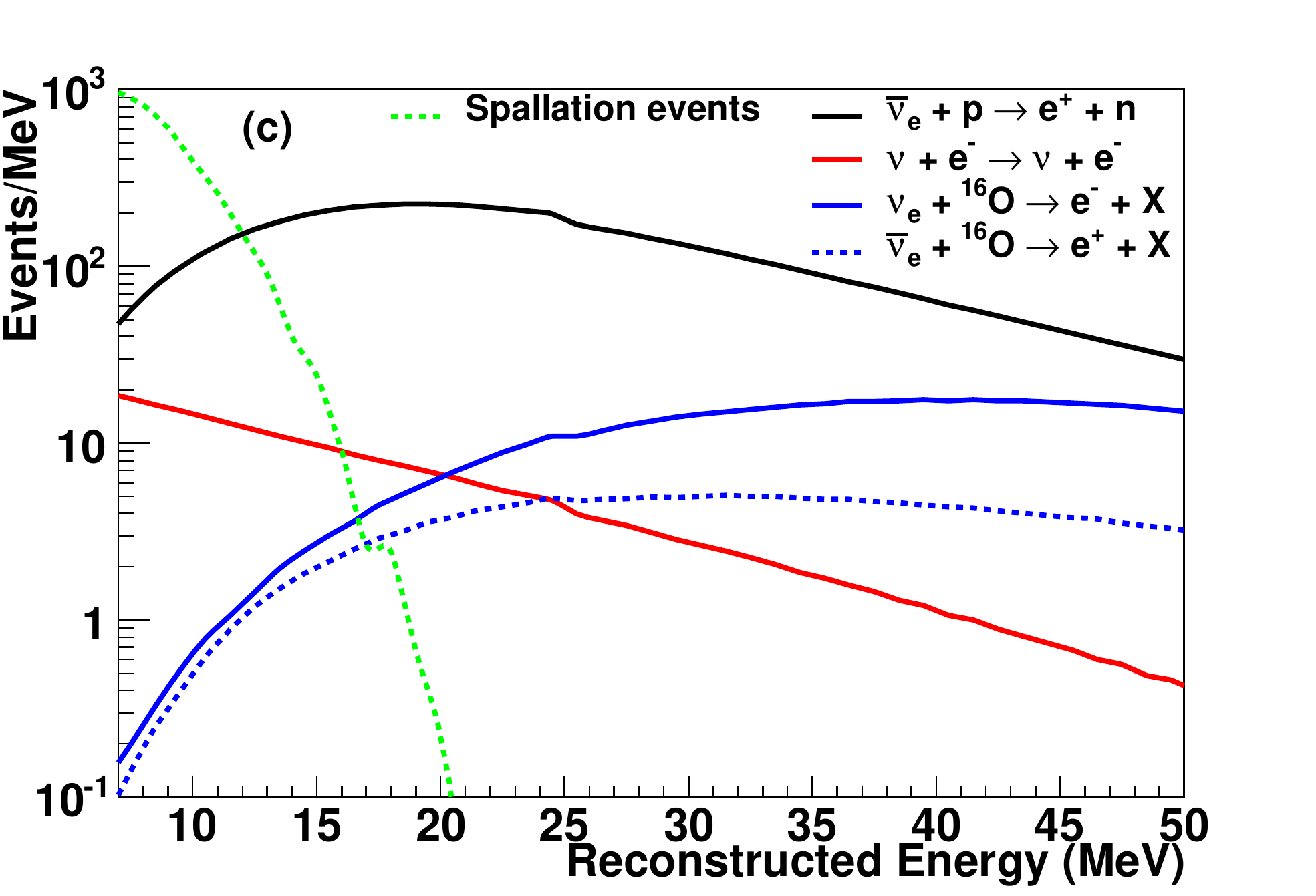}
\includegraphics[width=65mm,clip]{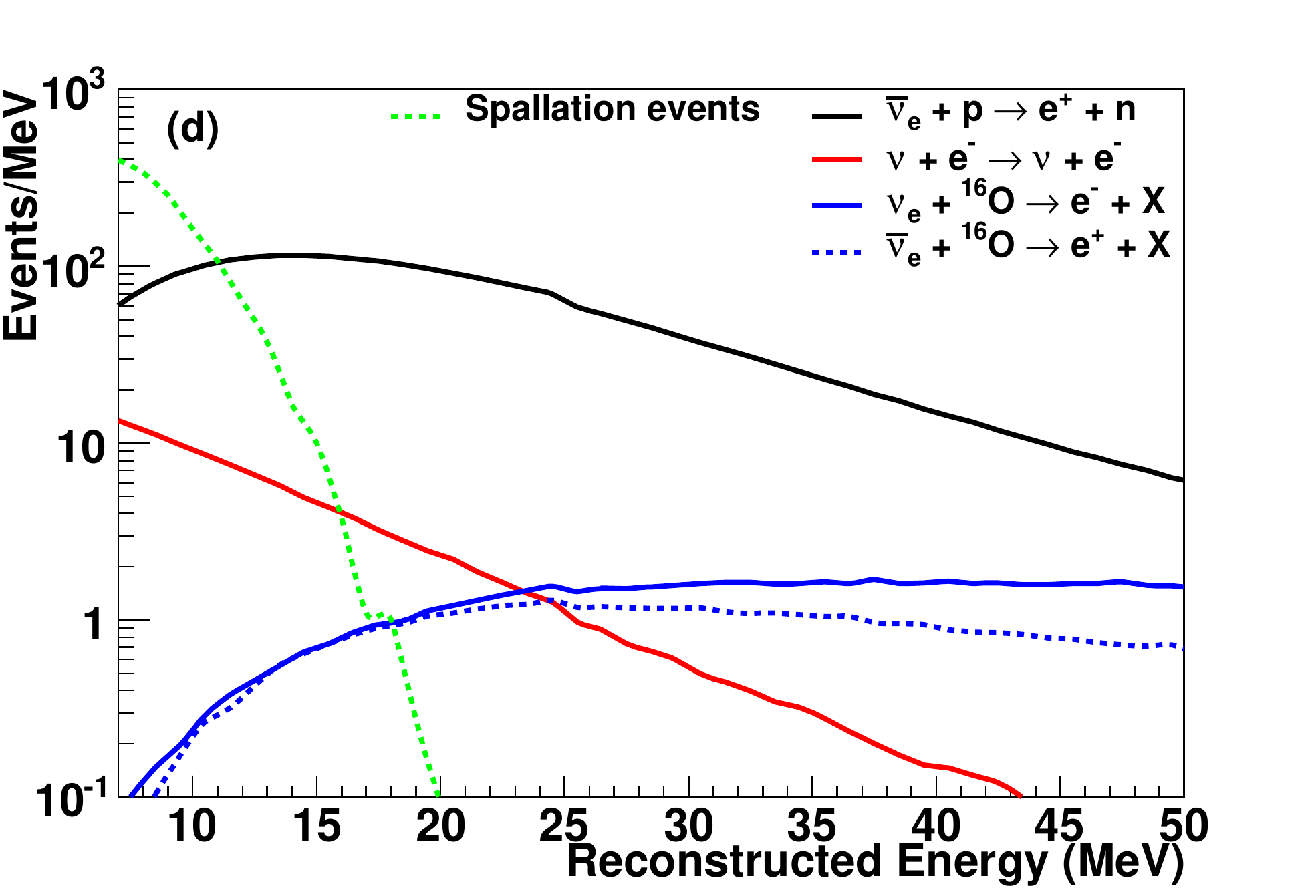}
\includegraphics[width=65mm,clip]{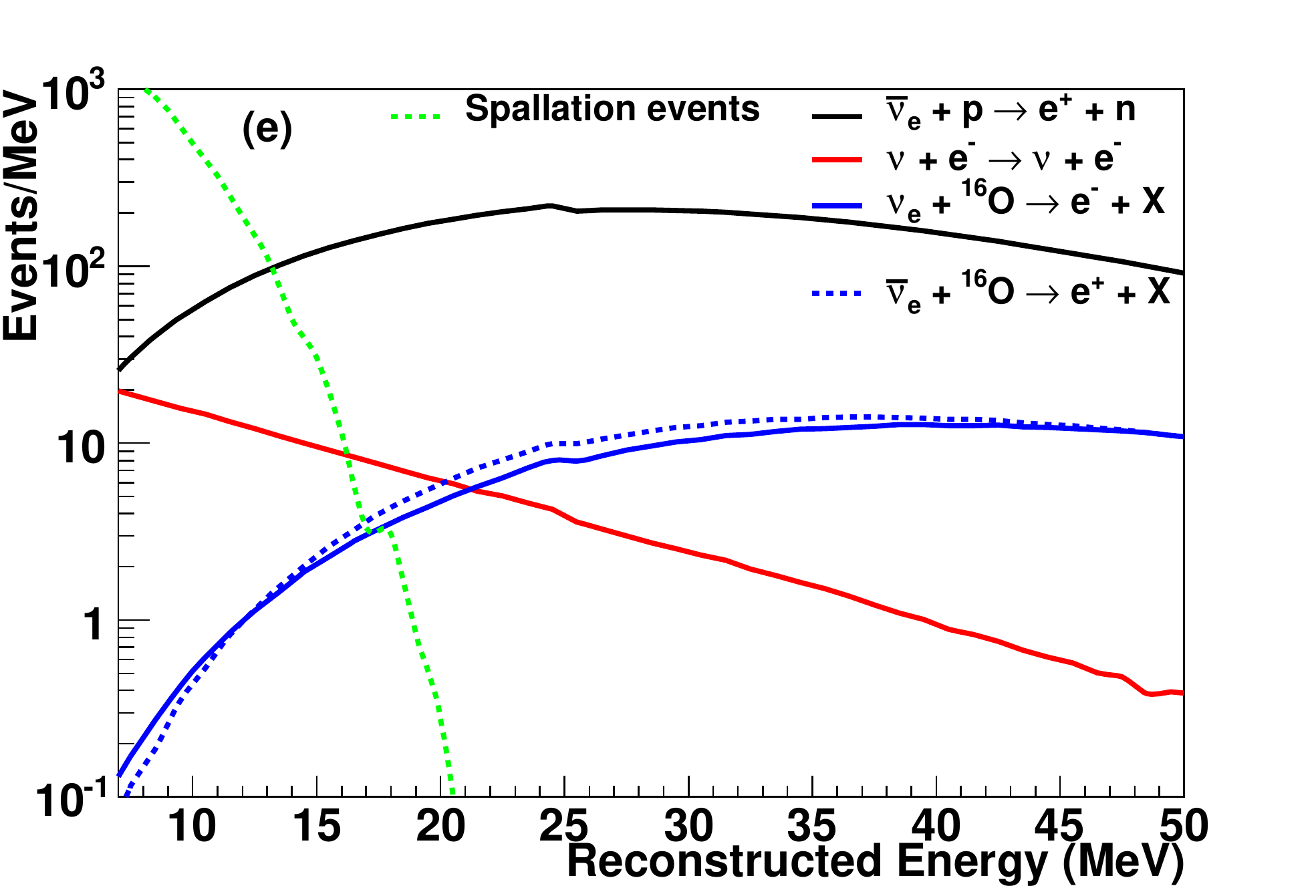}
\includegraphics[width=65mm,clip]{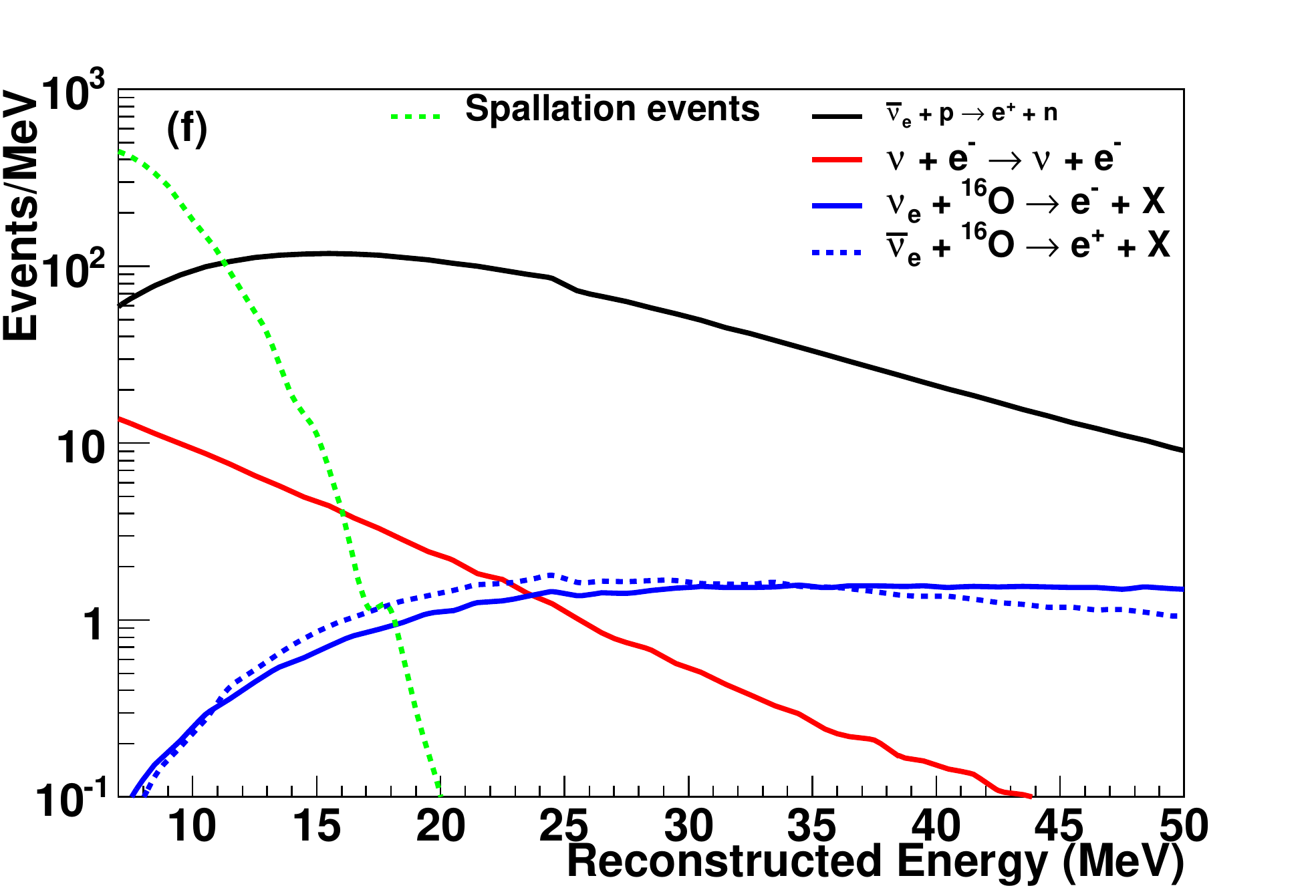}
\end{center}
\vspace{-5mm}
\caption{
Reconstructed energy distributions at SK for (a) the Wilson model
and (b) the NK1 model at 10~kpc for the four neutrino interaction channels
without neutrino oscillation, (c) the Wilson model and (d) the NK1 model
with the NH hypothesis for neutrino oscillations, and
(e) the Wilson model and (f) the NK1 model 
with the IH hypothesis for neutrino oscillations.
%\textcolor{blue}{
The green dotted line is the energy spectrum of the spallation candidates
found in the silent warnings.
The spallation event histogram is normalized to the number of 
SN MC entries.%}
}
\label{fig:sk_vis_ene}
\end{figure}

We generate MC samples for the three SN models for the three
neutrino oscillation hypotheses: no oscillation, normal hierarchy (NH)
and inverted hierarchy (IH).
% at a distance of 10~kpc.
Table ~\ref{table:expected-events} shows the expected numbers of events 
of the three SN models at SK
in the 22.5-kton fiducial volume with the total energy threshold of 7~MeV,
obtained by averaging the 3,000 MC ensembles at the distance of 10~kpc.
Figure~\ref{fig:sk_vis_ene} shows the reconstructed energy distributions at SK
for the Wilson and NK1 models with and without neutrino oscillations
to display the effect of the neutrino oscillations.
%\textcolor{blue}{
Figure~\ref{fig:sk_vis_ene} also shows the energy spectrum of the spallation
events found in the silent warnings described in Sec.~\ref{sec:monitor-system}.%}

%We show the energy spectra for NH hypothesis with neutrino oscillations,
%as the difference of the spectra between NH and IH hypotheses is not large.

\begin{table}
\begin{center}
\caption{Numbers of expected events at SK in the 22.5-kton fiducial volume with the 7~MeV 
total energy threshold for a SN burst with a distance of 10~kpc.
We estimated these numbers using SK MC: we generate 3,000 ensembles of the MC samples,
reconstructed the events with the SK standard reconstruction tool, applied the
selection criteria, and then calculated the average numbers.}
  \begin{tabular}{cccccccccc}
  \hline\hline
   & \multicolumn{3}{c}{ Wilson } & \multicolumn{3}{c}{NK1} & \multicolumn{3}{c}{NK2} \\
   &  no osc. &  NH  &  IH & no osc. & NH & IH & no osc. & NH & IH \\
   \hline
   $\bar{\nu}_e + p \to e^+ + n$ & 4923 & 5667 & 7587 & 2076 & 2399 & 2745 & 1878 & 2252 & 2652 \\
   $\nu_e + e^- \to \nu_e + e^-$ & 74   & 130  & 114  & 43   & 56   & 56 & 39 & 54 & 54 \\
   $\bar{\nu}_e + e^- \to \bar{\nu}_e + e^-$ & 25 & 29 & 37 & 10 & 12 & 14 & 9 & 11 & 13\\
   $\nu_x + e^- \to \nu_x + e^- $ & 41 & 33 & 34 & 17 & 19 & 18 & 17 & 17 & 17\\
   $\bar{\nu}_x + e^- \to \bar{\nu}_x + e^- $ & 34 & 33 & 29 & 14 & 14 & 14 & 13 & 13 & 14\\
   $\nu_e + ^{16}{\rm O} \to e^- + X$ & 8 & 662 & 479 & 22 & 78 & 74 & 16 & 72 & 68 \\
   $\bar{\nu}_e + ^{16}{\rm O} \to e^+ + X $ & 64 & 196 & 531 & 27 & 48 & 70 & 20 & 41 & 64\\
   \hline
   total & 5169 & 6750 & 8811 & 2209 & 2626 & 2991 & 1992 & 2460 & 2882 \\
   \hline\hline
   \end{tabular} 
\label{table:expected-events}
\end{center}
\end{table}

\subsection{The monitor system}
\label{sec:monitor-system}

\begin{figure}[h]
\begin{center}
\includegraphics[width=63mm,clip]{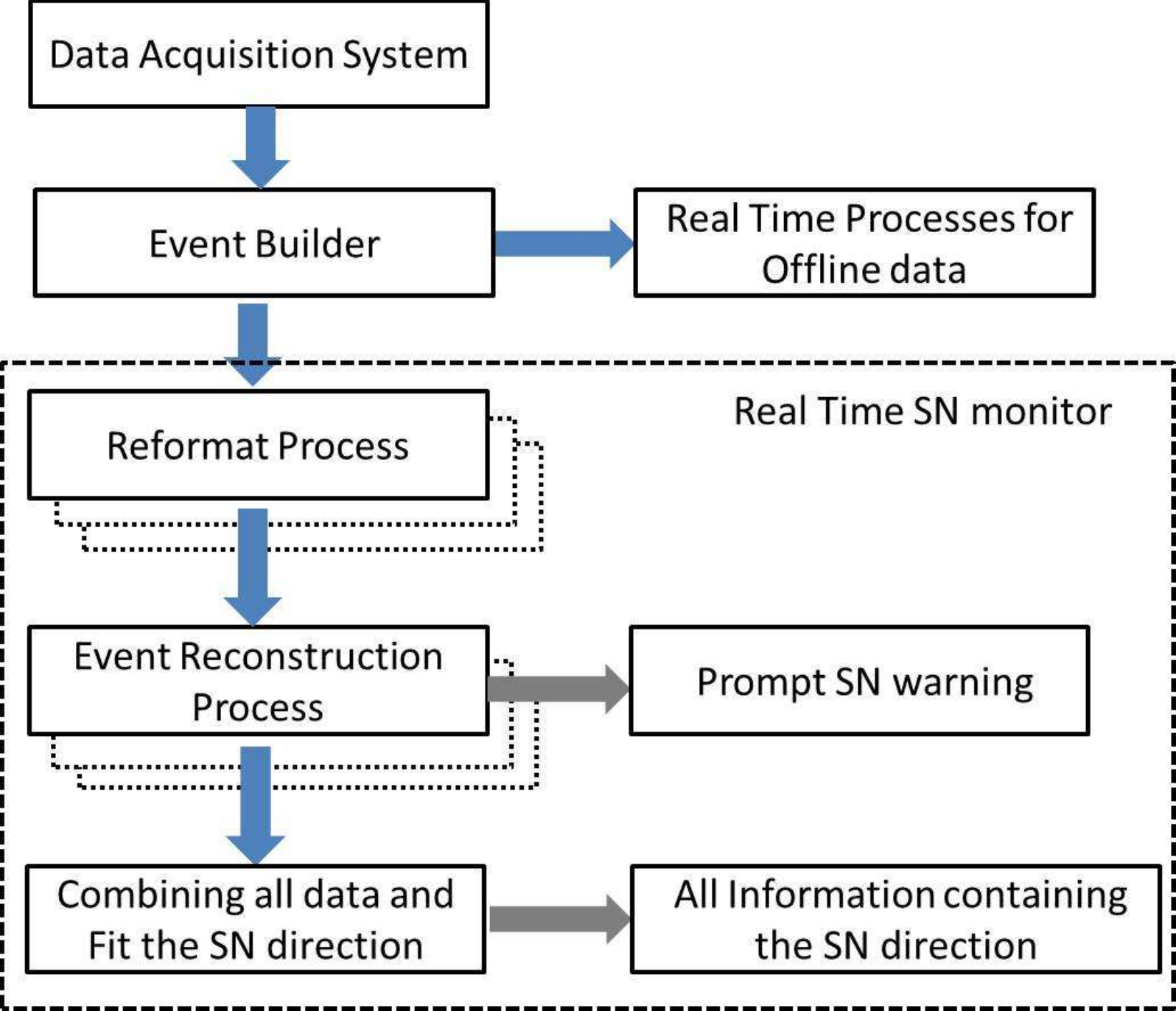}
\end{center}
\vspace{-5mm}
\caption{
Block diagram of the real time SN neutrino burst monitor.
The arrows show paths of the data flow, and boxes indicate
processes.
Details are explained in the text.
}
\label{fig:snwatch-system}
\end{figure}

In this section, we will describe details of the SN neutrino burst monitor system.
Figure~\ref{fig:snwatch-system} shows a flow diagram of the system.
The SK data collected by the data acquisition system are sent to the event builder.
At the event builder, the event data are packed and stored in a data file which we call a sub-run file.
Each sub-run file contains about one minute of event data.
The sub-run files are sent to both the offline process and the SN burst monitor.
In the offline process, the data files are converted to an offline data format that is used
for various physics analyses and detector calibrations.
The SN monitor system is running on a single computer on which a control process 
operating continuously handles all the processes and the data files.

For each sub-run file sent to the SN monitor, 
two processes are automatically executed by the control process:
a reformat process (the first process) and 
an event reconstruction process (the second process).
The reformat process converts the online data format to the offline data format.
Using the offline format data,
the event reconstruction process reconstructs the vertex position, direction and energy
for each event.
%\textcolor{blue}{
It takes about two minutes to finish the reformat and event reconstruction
for one sub-run file.%}
Events with total energy greater than 7~MeV and vertex position within the 22.5-kton
fiducial volume in SK are selected.
%\textcolor{blue}{
We remove cosmic ray muons and their subsequent decay electron events.%}
%where the fiducial volume is defined as the volume
%whose surface is located 2~m inside from the surface of the SK inner detector volume %(32.5~kton).
After the reconstruction of each selected event, a 20-second time window is opened backwards in time from the event, and the number of selected events 
in the window ($N_{\rm cluster}$) is counted.
%\textcolor{red}{
If there is a sub-run file boundary, 
the time window extends to the previous sub-run file.
%}

We also compute a variable $D$ that characterizes the vertex distribution.
The variable $D$ identifies the dimension of the vertex distribution and is an integer
number from 0 to 3, corresponding to point-, line-, plane- and volume-like distributions, respectively.
The variable $D$ is determined by comparing $\chi^2$ values obtained from 
the lengths of the major and minor axes
that correspond to the eigenvalues of a correlation matrix of the vertex distribution.  
The correlation matrix is a $3\times 3$ matrix whose elements are defined as
$\langle(x_i -\langle x_i \rangle)(x_j - \langle x_j \rangle)\rangle$,
where $i,~j = 1,~2,~3$ identify the vertex position axes and $\langle x \rangle$ is
the mean value of a variable $x$.
We construct a $\chi^2 = \sum_{n=1}^{N_{\rm cluster}}|\vec{d}_n - \vec{d}(\vec{s})|^2$,
where $\vec{d}_n$ is the $n-$th event vertex position and $\vec{d}(\vec{s})$
is a position closest to $\vec{d}_n$ on either a point, a line or a plane
with parameters $\vec{s}$ that determine the geometry of the three cases.
The three eigenvalues $\lambda_i$ ($i=1,~2,~3$, and $\lambda_1 \leq \lambda_2 \leq \lambda_3$) are used to construct the minimum $\chi^2$ values
that are $(\lambda_1 + \lambda_2 + \lambda_3)/3$,
$(\lambda_1 + \lambda_2)/2$ and $\lambda_1$,
computed by assuming 
the vertex distribution is point-, line- and plane-like, respectively,
with the condition of $\partial \chi^2 / \partial \vec{s}=0$.
The comparison of the $\chi^2$ values to determine a $D$ value 
is tuned using MC simulations
so that the calculated $D$ value reproduces the input one.
%The value $R_{\rm mean}$ is the mean distance between two vertices for all the combinations.
%A concentrated (diffused) vertex distribution yields a small (large) $R_{\rm mean}$ value.
In case of a real SN burst, the vertex distribution should be uniform in SK, and we would have $D=3$,
depending on the number of burst events, which is confirmed by a simulation.
In contrast, for the case of a background burst mainly originating from spallation events,
the vertex positions distribute along the parent muon tracks, and we would have $D=2$, 1 or 0,
where the spallation events are the radioactivities
created 
%by the collisions on oxygen nuclei by high-energy cosmic ray muons.
by both high-energy cosmic ray muons and by constituents of
the resulting hadronic showers.
When the process finds $N_{\rm cluster} \geq 60$ events and $D=3$,
it generates a prompt SN burst warning which initiates phone-callings and emails sent 
to experts in the SK collaboration 
%\textcolor{red}{
within a few minutes after the SN burst occurs.
%}
We call such a warning a ``golden'' warning.
Subsequent to a golden warning, the experts start a meeting in order to make a world-wide announcement within one hour.
The threshold of $N_{\rm cluster}$ is determined so that
we would have 100\% SN detection efficiency at the Large Magellanic Cloud (LMC)
assuming the three SN models described in Section~\ref{sec:SN-MC}.
%For the case that a SN explosion happens at the center of our galaxy,
%a distance of about 10~kpc from the earth,
%we expect $5,000 \sim 8,000$ events in the fiducial volume 
%with total energy greater than 7~MeV,
%assuming the Wilson model~\cite{wilson-model},
%depending on the neutrino mass hierarchy hypothesis.
%The expected number of accidental background events satisfying the event selection %events originating 
%from spallation during the neutrino burst is 0.7 events per 10 seconds, 
%estimated by counting the number of events with energies greater than 7~MeV
%in the fiducial volume using a data sample of SK.
%This estimation is valid since with the selection condition
%there is no contamination from known radioactivities
%other than the spallation products. 
%Since the reconstruction process treats only a single sub-run file, 
%the SN neutrino burst events may be separated to multiple sub-run files.
%To collect all the burst events, the third process is executed 
%by the control process in Fig.~\ref{fig:snwatch-system}.

The third process in Fig.~\ref{fig:snwatch-system} combines all the sub-run data and
determines the SN direction by a fit.
All the SN burst event information is summarized and 
sent to the experts by e-mail,
which is also used as the input to the discussions.
%The prompt warning is generated about 2~minutes after the SN neutrino burst,
%which is measured by flashing a LED placed inside the SK that generates an event burst
%with the mean energy of 27~MeV.
%The second warning is issued about 6~minutes later after the burst
%that has 8,000 events.
%The delay is caused by the time spent by the reconstruction of all the events and
%a SN direction fit.
%The prompt warning ensures so that the experts can be ready to join the meeting as soon as possible,
%independent of the time delay of the second warning. 
Following these discussions the announcement containing the information
about the number of observed neutrinos,
%The announcement containing the information about the number of observed neutrinos,
the burst time duration, 
the 
%\textcolor{red}{
universal time
%}
the burst happens 
and the estimated direction of the SN 
%\textcolor{red}{
in the equatorial coordinate system
%}
is broadcast to the ATEL~\cite{atel}, GCN~\cite{gcn}, IAU-CBAT~\cite{cbat}
and SNEWS~\cite{snews}.
%\textcolor{red}{
The universal time is determined using 1 pps (pulse per second) signals from 
the global positioning system and a local time clock system consisting of
a commercial rubidium clock~\cite{t2k-gps}. 
%}
No golden prompt warning
has been sent so far.
%has been alerted so far for 16 years of the SK operation.

When the SN burst has less than 60 events, the golden warning will not be generated.
Instead we set another threshold.
The threshold for generating the warning is determined so that the backgrounds are suppressed:
we set the threshold of $N_{\rm cluster} \geq 25$ and require $D=3$.
%Two thresholds are provided.
%One threshold is set so that
%a warning is issued when $N_{\rm cluster}>25$ and $R_{\rm mean}> 900~(750)$~cm for
%$N_{\rm cluster}<40$ ($N_{\rm cluster}>40$).
The warning generated with this condition is called a normal warning.
The normal warning is sent to the experts only by e-mail, 
without phone-calls, and to SNEWS.
Conveners among the experts check the event cluster found by this warning and
make a decision about whether to have a meeting for the announcement.
The normal warning threshold is set so that
we would have 100\% SN detection efficiency at the Small Magellanic Cloud (SMC)
assuming the three SN models.
The details of the detection efficiency will be described in 
Section \ref{sec:performance-of-snwatch}.
The reason to provide the normal warning is to avoid
any fake warnings caused by unexpected software and hardware troubles.
We have had no normal warning so far.
%In fact, we had a few fake warnings that satisfy the normal warning condition but not
%the real SN.
%Those fake warnings were found to be fake immediately 
%after the warning issue by checking various plots provided by the 
%third process in Fig.~\ref{fig:snwatch-system}.
%The normal warning has been issued once per three months due mostly to the spallation events
%caused by multiple high energy cosmic ray muons (a muon bundle) which make 
%scatter vertex distributions having large $R_{\rm mean}$ values.

%\textcolor{blue}{
%In summary, the SN monitor performs the reformat of the data and
%the reconstruction of the vertex
%position, the direction and the energy for events in a sub-run file
%in a processing time of a few minutes,
%and searches for an event burst with a time window of 20 seconds for events
%in the 22.5-kton fiducial volume with the energy greater than 7~MeV.
%When a warning is issued, the experts are collected to have a meeting for
%a public announcement within an hour.}
In summary, the SN monitor reformats the data and determines the vertex position, direction, and energy of events within a few minutes of the data being collected.  It then searches for bursts of events - a cluster occurring within 20 seconds - with energies above 7 MeV and whose vertices fall within SK's 22.5-kton fiducial volume. Within an hour of a warning being issued, the experts gather and hold a meeting to determine the appropriate public announcement to make, if any, based on the quality and nature of the detected burst.%}

We also provide a lower threshold such that we require
%more than 7 events in 0.5~seconds,
%more than 8 events in 2.0 seconds and 
more than 13 events in 10 seconds.
We call a warning generated with this condition a ``silent'' warning.
The conditions were tuned so that we would have a few warnings per day from spallation events.
%In fact, the silent warning happens about two times per day on average.
The silent warnings are sent to only a few experts
of the monitor system operation and detector condition,
and are not used as the fast alert for a SN burst.
%The silent warning rate per day for the last 3 years is shown in Fig.~\ref{fig:alarm-rate}.
%We find the system has a stable operation.

%\begin{figure}[h]
%\begin{center}
%\includegraphics[width=63mm,clip]{alarm_rate}
%\end{center}
%\vspace{-5mm}
%\caption{
%Silent alarm rate per day during the last 3 years.
%We have 2.7 silent alarms per day on average by spallation events.
%}
%\label{fig:alarm-rate}
%\end{figure}

%
% here we put Rmean vs Ncluster and D distributions to demonstrate the selection criteria
%

%Figure~\ref{fig:rmean-ncluster} shows a scatter plot of $R_{\rm mean}$ and $N_{\rm cluster}$
%of event clusters for SN MC and data which include the normal and silent warnings.

\subsection{Performance study of the SN burst monitor with simulations}
\label{sec:performance-of-snwatch}

\begin{figure}[h]
\begin{center}
\includegraphics[width=65mm,clip]{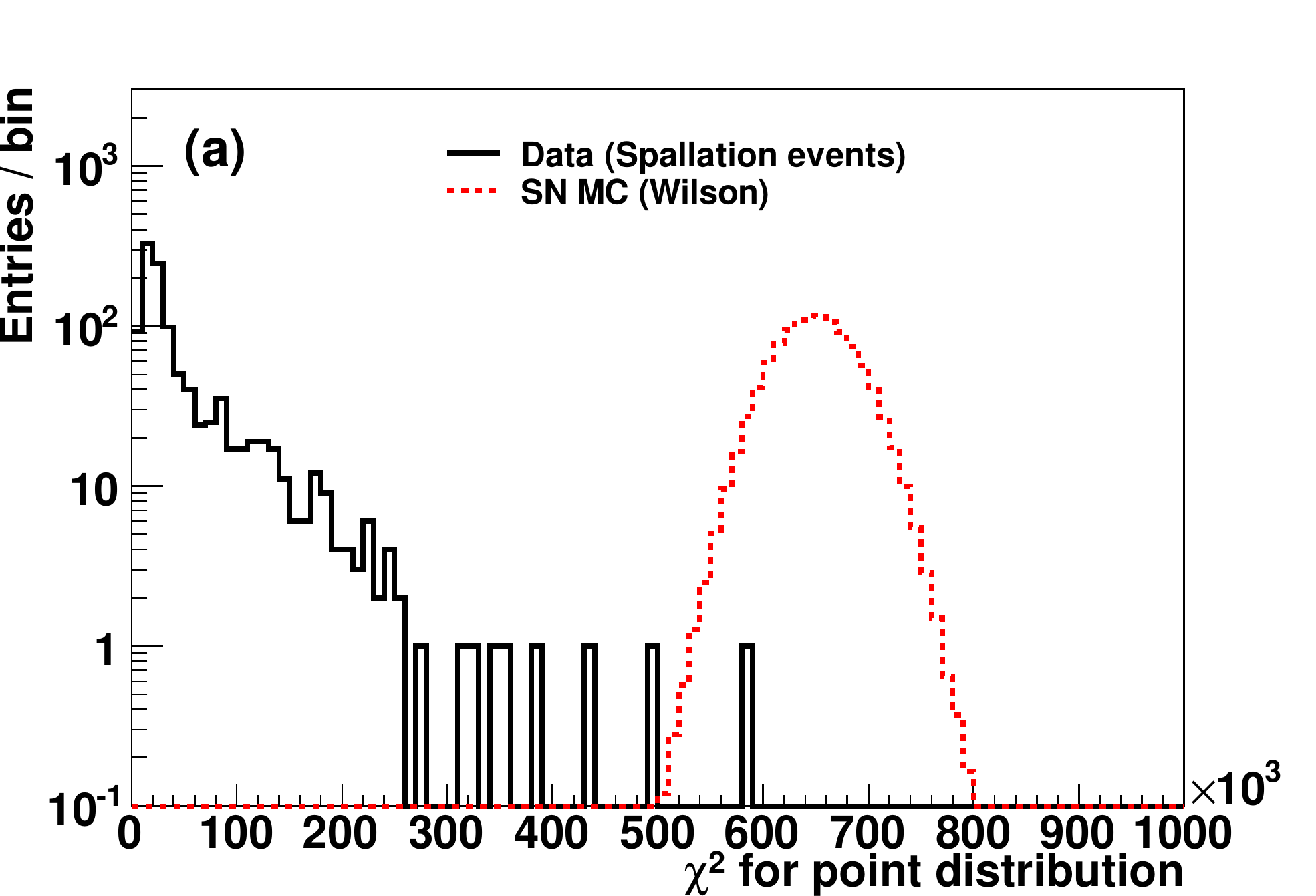}
\includegraphics[width=65mm,clip]{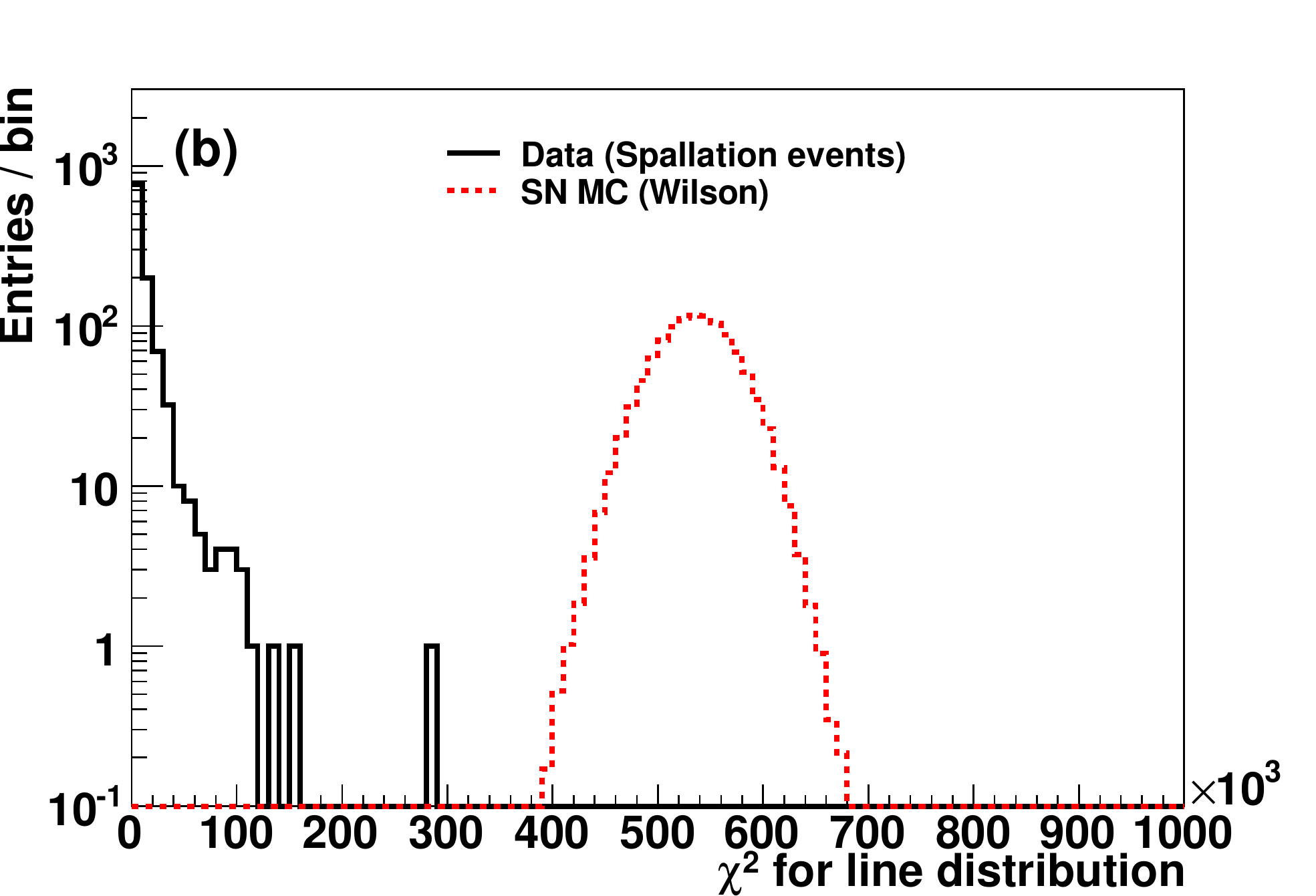}
\includegraphics[width=65mm,clip]{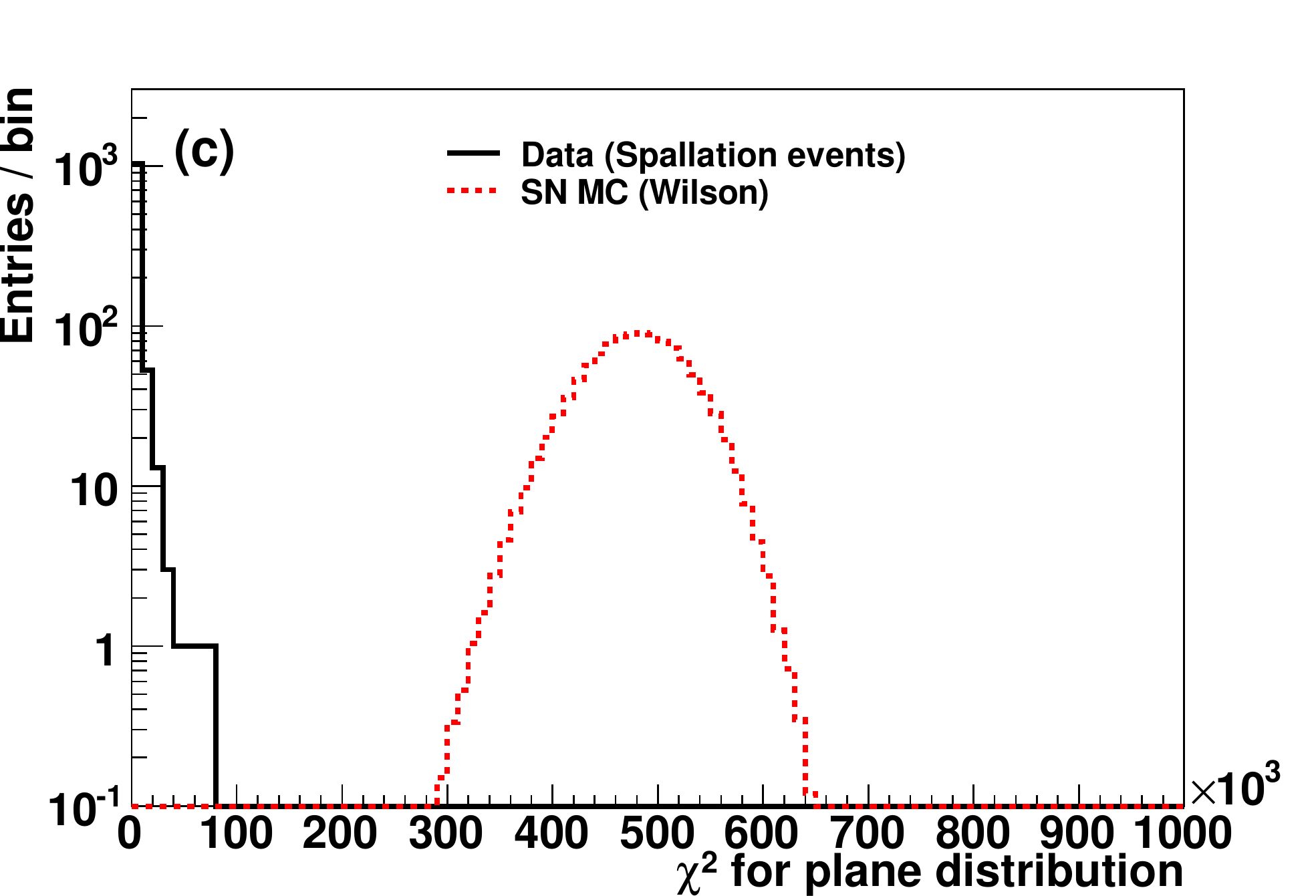}
\includegraphics[width=65mm,clip]{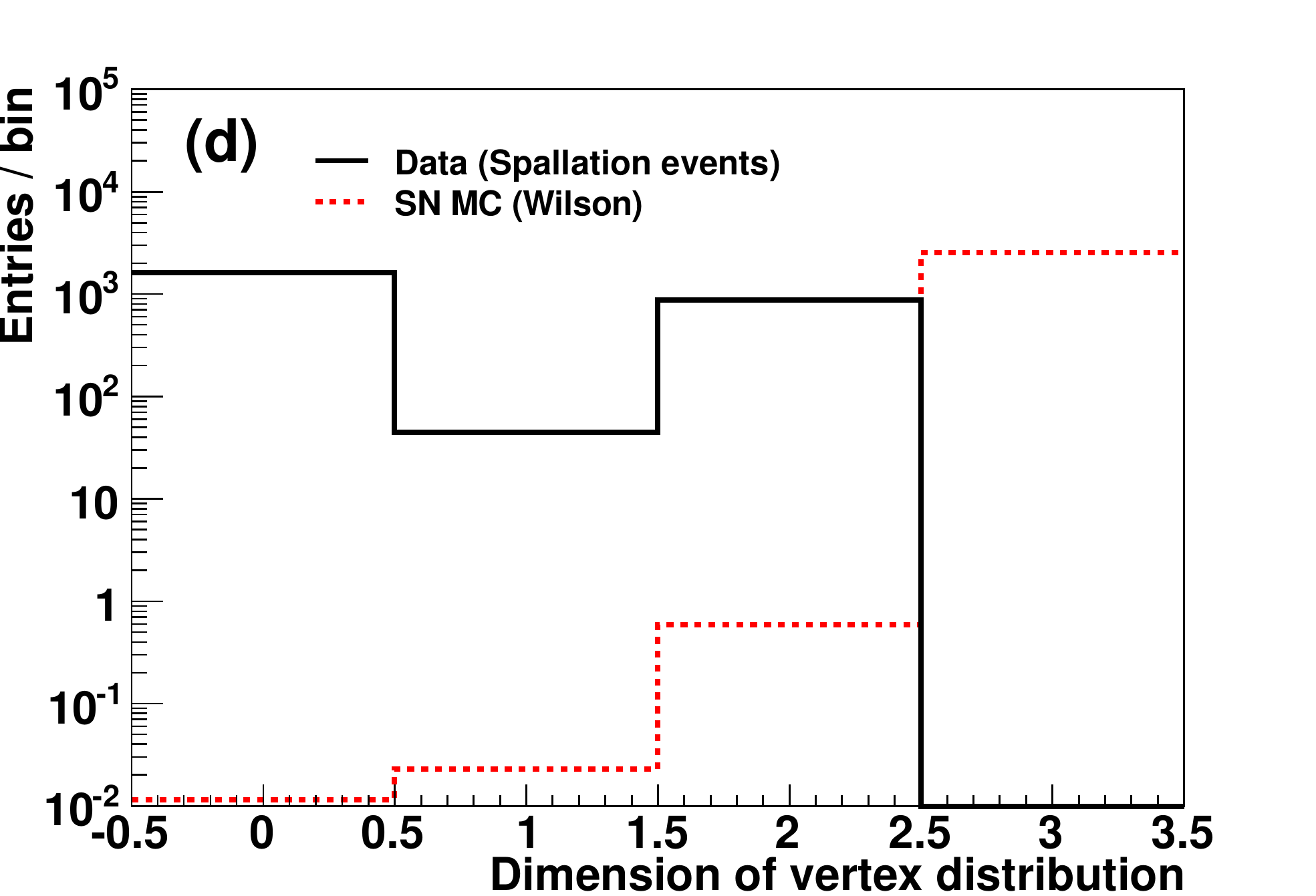}
\end{center}
\vspace{-5mm}
\caption{
%\textcolor{blue}{
Minimum $\chi^2$ value distributions assuming (a) a point distribution,
(b) a line distribution and
(c) a plane distribution, and%}
(d) distribution of $D$ (dimension of vertex distribution)
for SN MC (red broken line)
and spallation data (black solid line) found as silent warnings.
For SN MC clusters, we plot the distributions for the MC samples with
$60 \leq N_{\rm cluster} \leq 100$.
The histogram of SN MC is normalized to the number of data entries.
}
\label{fig:dimension}
\end{figure}

%\textcolor{blue}{
Figure~\ref{fig:dimension} shows the minimum $\chi^2$ distributions
for the three geometrical assumptions and
$D$ distributions for SN MC and spallation data
triggered by the silent warnings.%}
The spallation event cluster is identified from the vertex, energy and time distributions,
i.e., the vertex distribution is concentrated around the parent muon track,
the energies of the events have the typical spallation energy spectrum up to 20~MeV,
and the time distribution is an exponential decay consistent with the lifetimes
of the spallation products.
In the figure, we generate MC simulation samples in the range of $60 \leq N_{\rm cluster} \leq 100$
uniformly, and plot the distributions for the samples.
The probability to have $D\leq 2$ for SN MC with the $N_{\rm cluster}$ range is
$8\times 10^{-4}$. 
%\textcolor{red}{
No SN MC sample having $D \leq 2$ is found for 930,000 samples
with $100<N_{\rm cluster}<1,000$.
%}
For a normal warning condition, i.e., $25 \leq N_{\rm cluster}<60$, 
the probability to have $D \leq 2$ is 1.3\%.
This demonstrates that the variable $D$ can discriminate between the
SN-like clusters and spallation background clusters.

%\begin{figure}[h]
%\begin{center}
%\includegraphics[width=73mm,clip]{rmean_vs_mult}
%\end{center}
%\vspace{-5mm}
%\caption{
%Scatter plot of $R_{\rm mean}$ and $N_{\rm cluster}$ of event clusters for SN MC (black circule)
%and data (blue square) that includes normal and silent warnings caused by the 
%spallation events.
%The solid and broken lines show the selection criteria of the warnings of gold and normal,
%respectively.
%}
%\label{fig:rmean-ncluster}
%\end{figure}

\begin{figure}[h]
\begin{center}
\includegraphics[width=73mm,clip]{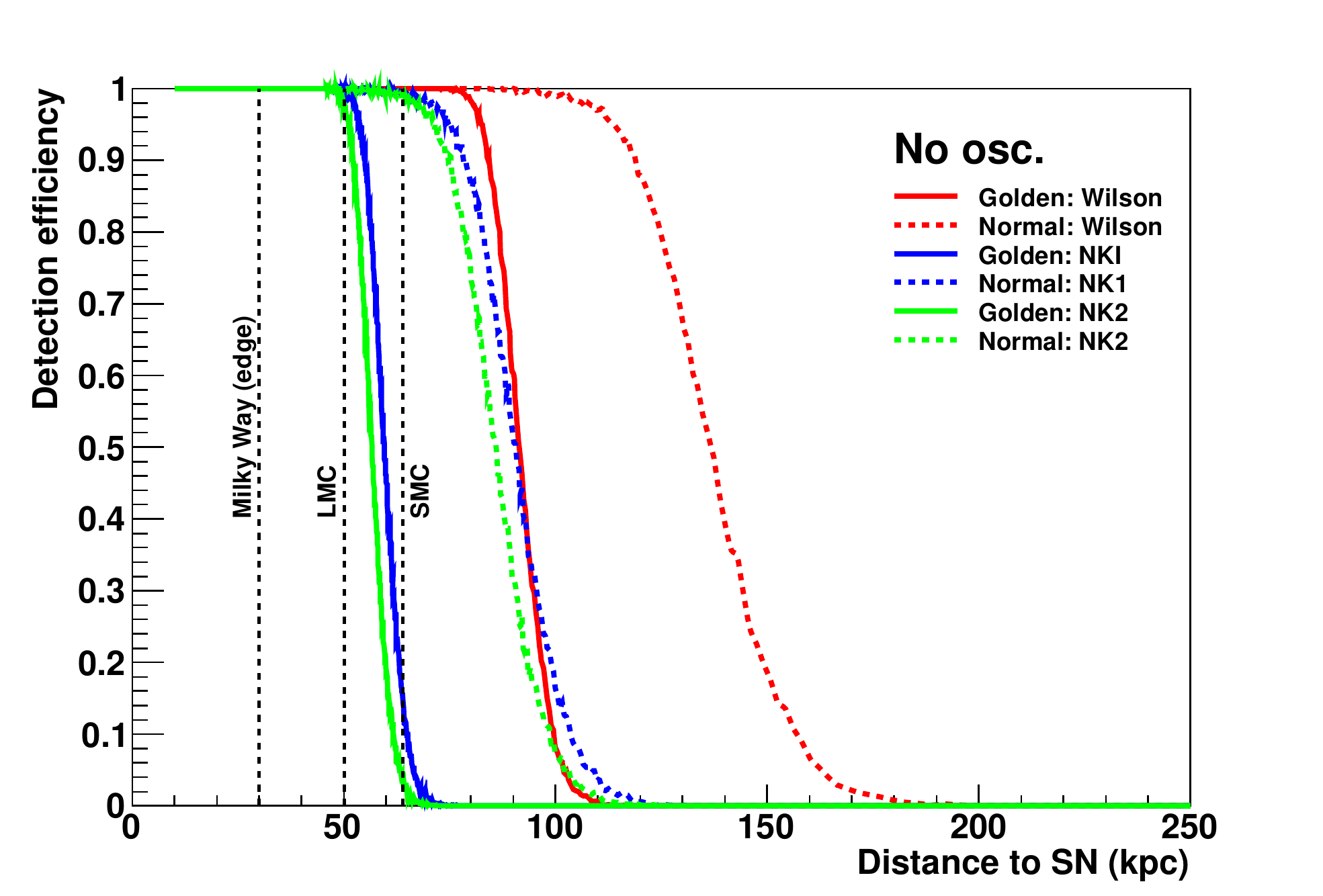}
\includegraphics[width=73mm,clip]{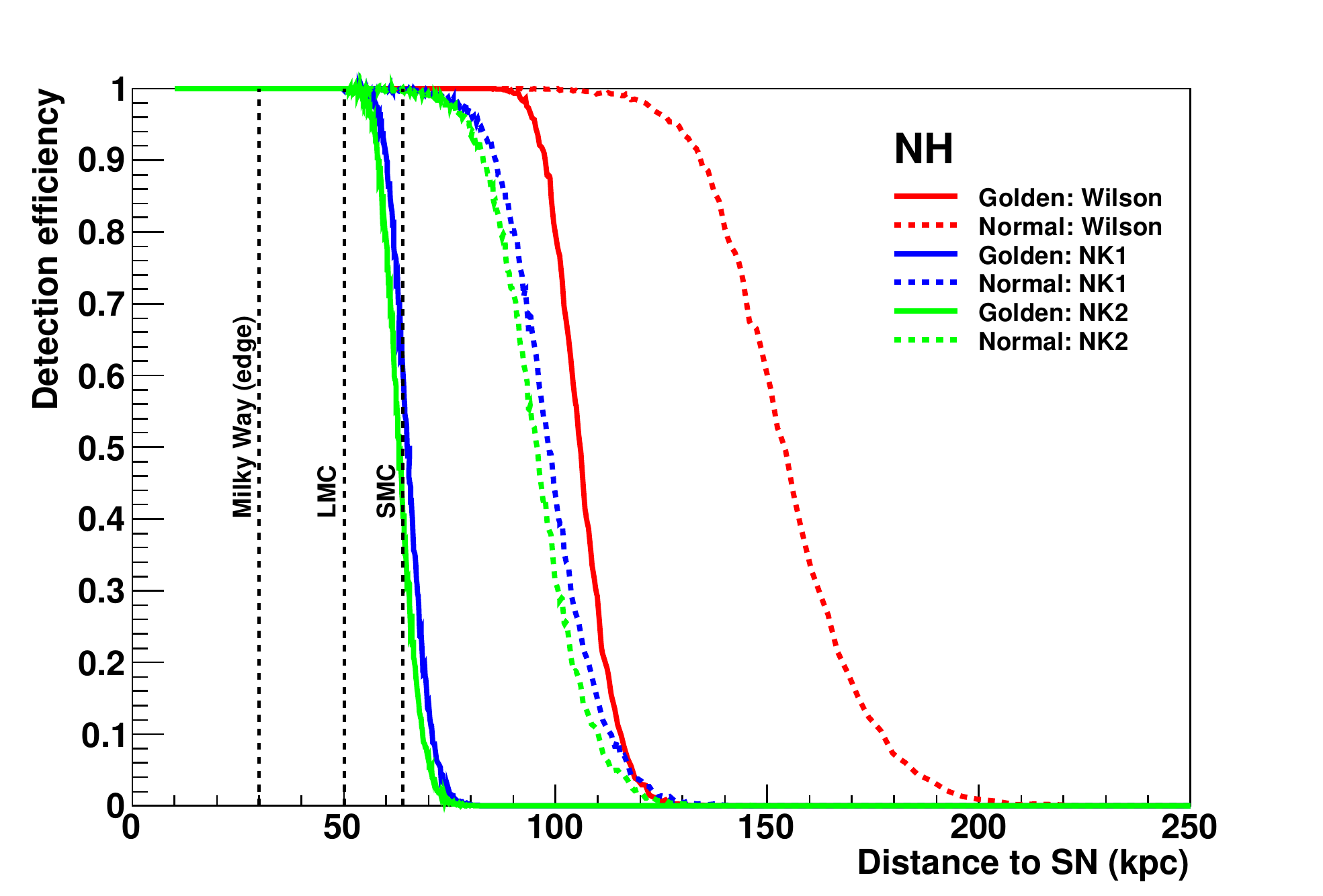}
\includegraphics[width=73mm,clip]{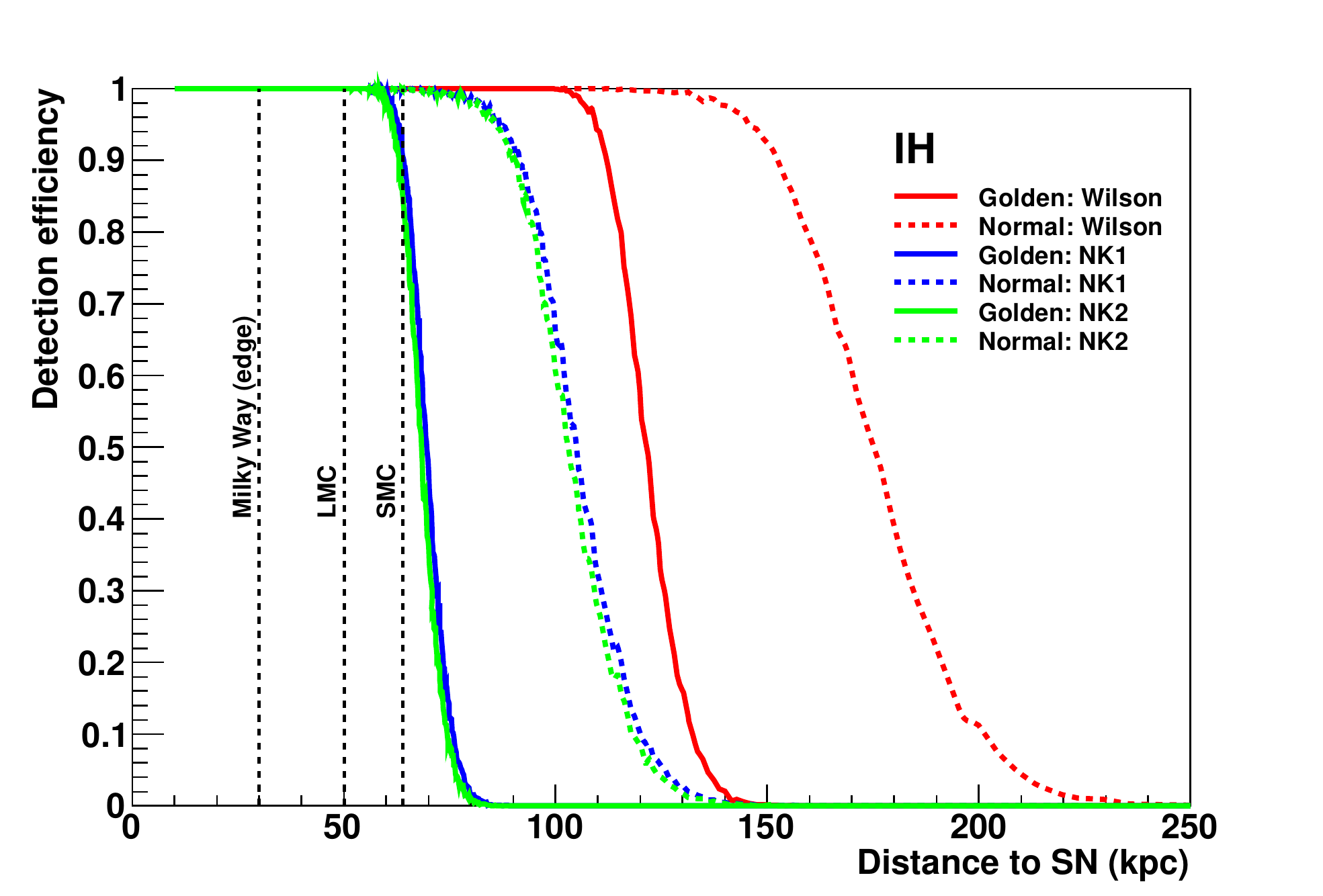}
\end{center}
\vspace{-5mm}
\caption{
Detection efficiency of a SN burst as a function of a distance for the normal
and golden warnings with the Wilson and Nakazato models.
From top to bottom, without neutrino oscillation,
with neutrino oscillations for the normal hierarchy and with
neutrino oscillations for the inverted hierarchy.
}
\label{fig:sn-eff}
\end{figure}

Figure~\ref{fig:sn-eff} shows the SN detection efficiency as a function of a
distance to a SN for the normal and golden warnings for the three SN models
without neutrino oscillation and with neutrino oscillations for
normal and inverted mass hierarchy hypotheses.  
It is found that the system has 100\% detection efficiency 
up to the LMC located at 50~kpc away for all three models
for the golden warning with the three hypotheses.
For the SNe at the SMC, about 64~kpc away,
the efficiency depends on the hypotheses for the Nakazato model,
and is 100\% for the Wilson model.
The normal warning has almost 100\% detection efficiency
for the three models.
The efficiency is basically determined by the number of inverse beta decay events.
The difference between detection efficiencies among the three hypotheses of the neutrino oscillations
is caused by the difference between the $\bar{\nu}_e$  energy spectra.
The average energy of $\bar{\nu}_e$ is smaller than that of $\bar{\nu}_x$
when those neutrinos are emitted from the neutrinosphere.
%If neutrino oscillations happen,
%\textcolor{red}{
With neutrino oscillations,
%}
$\bar{\nu}_x$ are converted to $\bar{\nu}_e$ and therefore
the average energy of $\bar{\nu}_e$ at SK increases,
resulting in a higher event rate of the inverse beta decays.
Due to this effect, the detection efficiency of SNe at the SMC increases
for the case of neutrino oscillations.

\subsection{Operation of the SN burst monitor}
\label{sec:operation}

%\textcolor{blue}{
We have operated the SN burst monitor system for about 20 years, 
since the beginning of SK data-taking in 1996. 
The SN burst selection criteria and operation scheme have been changed, updated, 
and improved throughout this period. 
The SN monitor system scheme described in this report came into service 
in April of 2013. 
Before that time, earlier versions of the monitor system had been running 
as one of the offline processes.%}

%\textcolor{blue}{
In Fig. 5 (a), we show the silent warning rates per 24 hours 
as a function of the elapsed days from Jan. 1st, 2010. 
The rate has trends as a consequence of -- and which track -- variations 
of the water transparency in SK. 
The energy scale used in the energy reconstruction program 
in the SN monitor process has been continually adjusted to 
compensate for these transparency fluctuations. 
Despite these fluctuations, the warning rate has been relatively 
stable over the last six years, with an average rate of 2.4 warnings per 24 hours.%}

%\textcolor{blue}{
The expected number of accidental background events satisfying 
the event selection is 0.121 events per 10 seconds 
with a root mean square of 0.007 events. 
We estimate this by counting the number of events in the SK fiducial volume with a total energy greater than 7 MeV for one day, 
and scale this number to a rate per 10 seconds. 
Figure 5 (b) shows the estimated averaged background event rate 
for a recent period of 434 days. 
The background events are considered to be spallation products, 
since there should be negligible contamination from known radioactivities 
other than spallation products given the applied energy threshold.%}

%\textcolor{blue}{
Figure 5 (c) shows the data processing time distribution for the silent warnings found. 
The average time to finish the processing is about 170 seconds; 
fluctuations are caused by the reconstruction process and the
condition of the network through 
which the data sample files are copied from the SK data acquisition system. 
The offline SN monitor that had run before April 2013 took about five minutes 
to finish the reformat and reconstruction processes, 
as the offline reconstruction program was tuned for physics analysis 
and calibration. 
We have optimized the reconstruction program for the online SN monitor 
to increase the processing speed without degrading its performance.%}

%\textcolor{blue}{
Figure 5 (d) shows the averaged monthly duty cycle of the SN monitor system 
over a recent 34-month period; the SN monitor operates with a duty cycle of 
about 97\%. The monitor searches for SN event bursts during normal SK running, 
but it does not operate during SK detector calibration runs, 
particularly during those calibration runs employing artificial sources 
that intentionally generate event bursts. 
Most of the 3\% loss of the SN monitor duty cycle comes from planned calibration.
Note that even when the SN monitor is off, SK still has a non-realtime capability 
to detect a SN burst during these calibration runs.  
This is achieved via dedicated offline analyses which remove likely source 
events based on their vertex positions and event timings.%}

\begin{figure}[h]
\begin{center}
\includegraphics[width=60mm,clip]{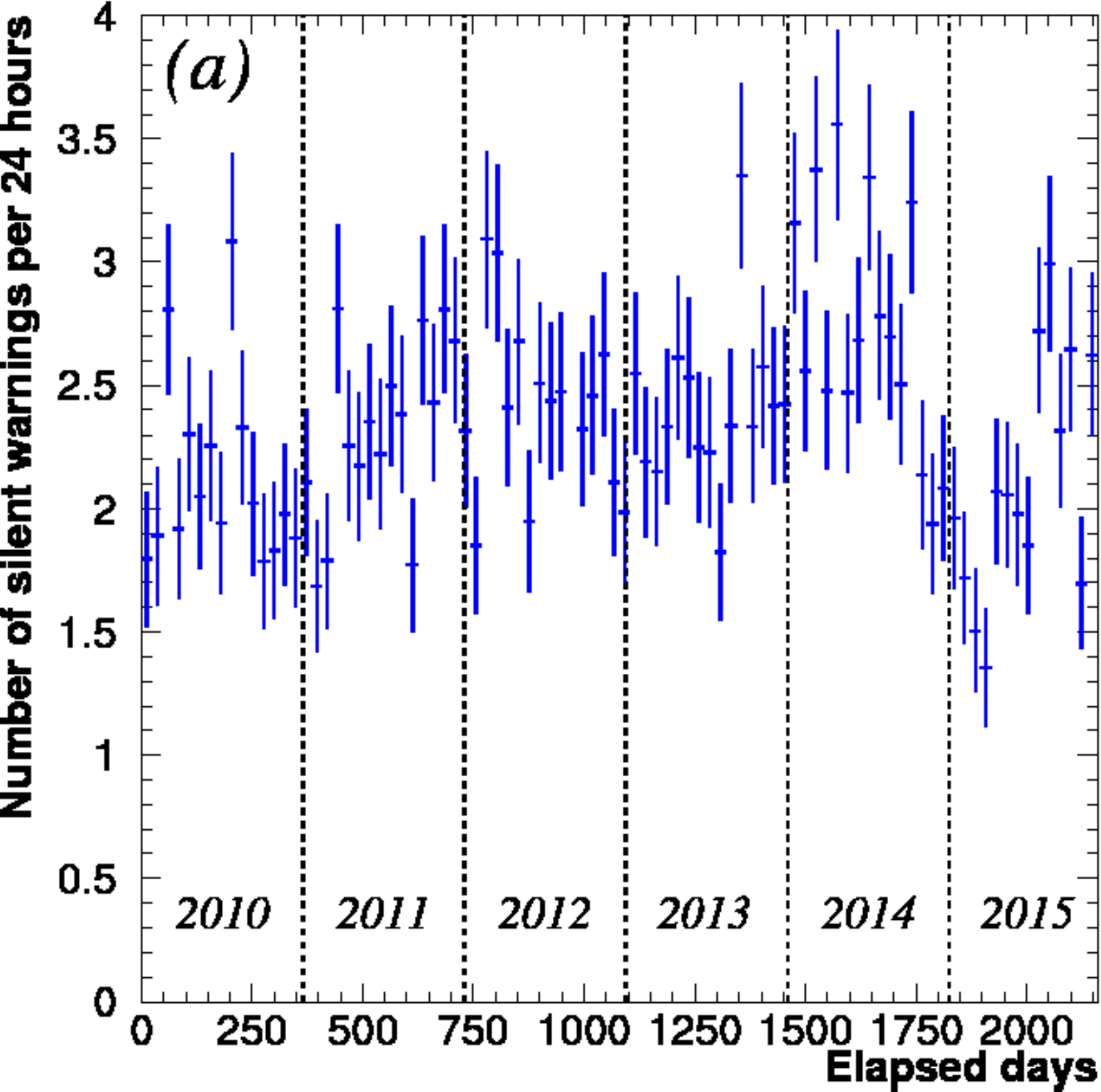}
\includegraphics[width=60mm,clip]{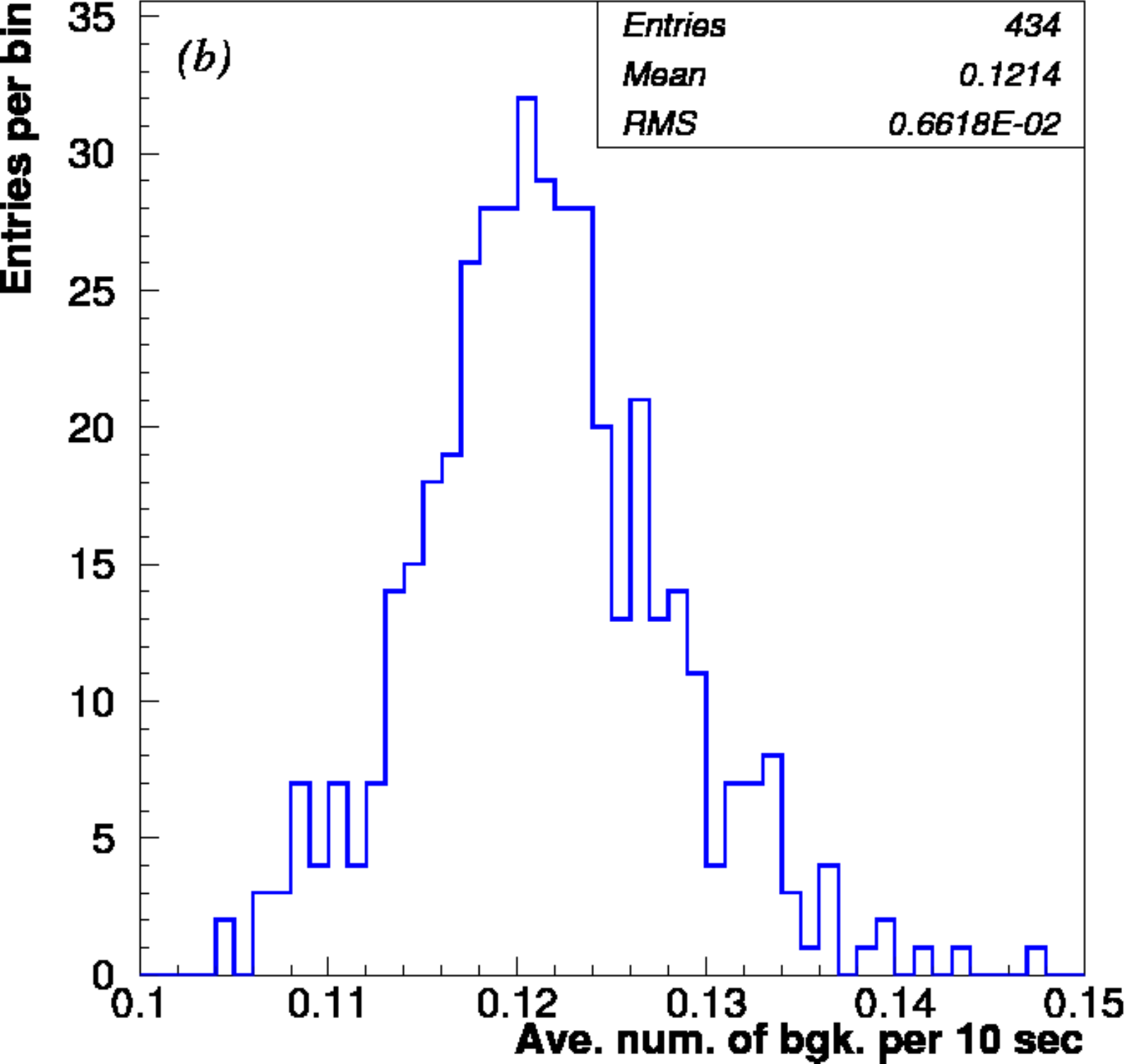}
\includegraphics[width=60mm,clip]{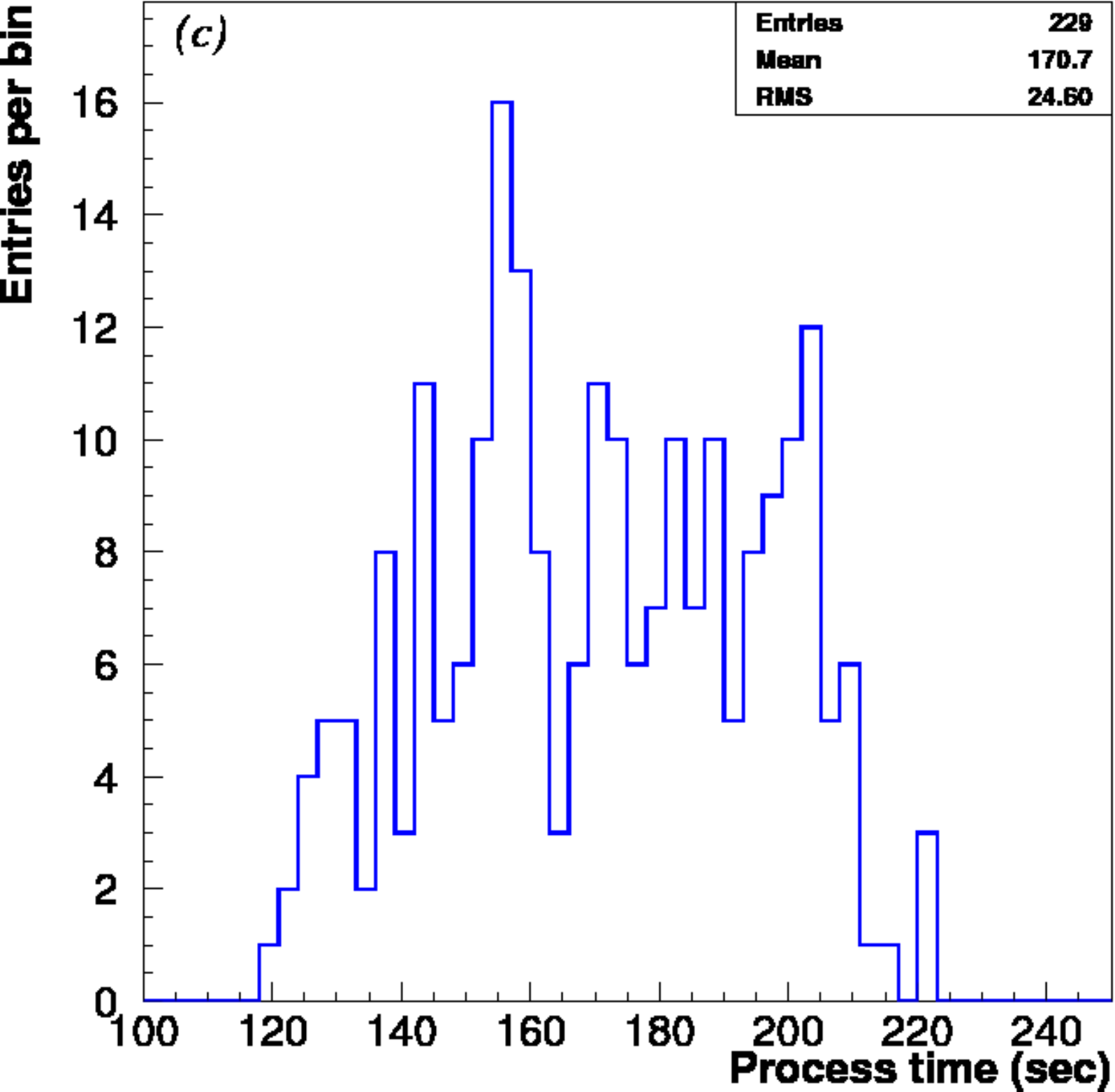}
\includegraphics[width=60mm,clip]{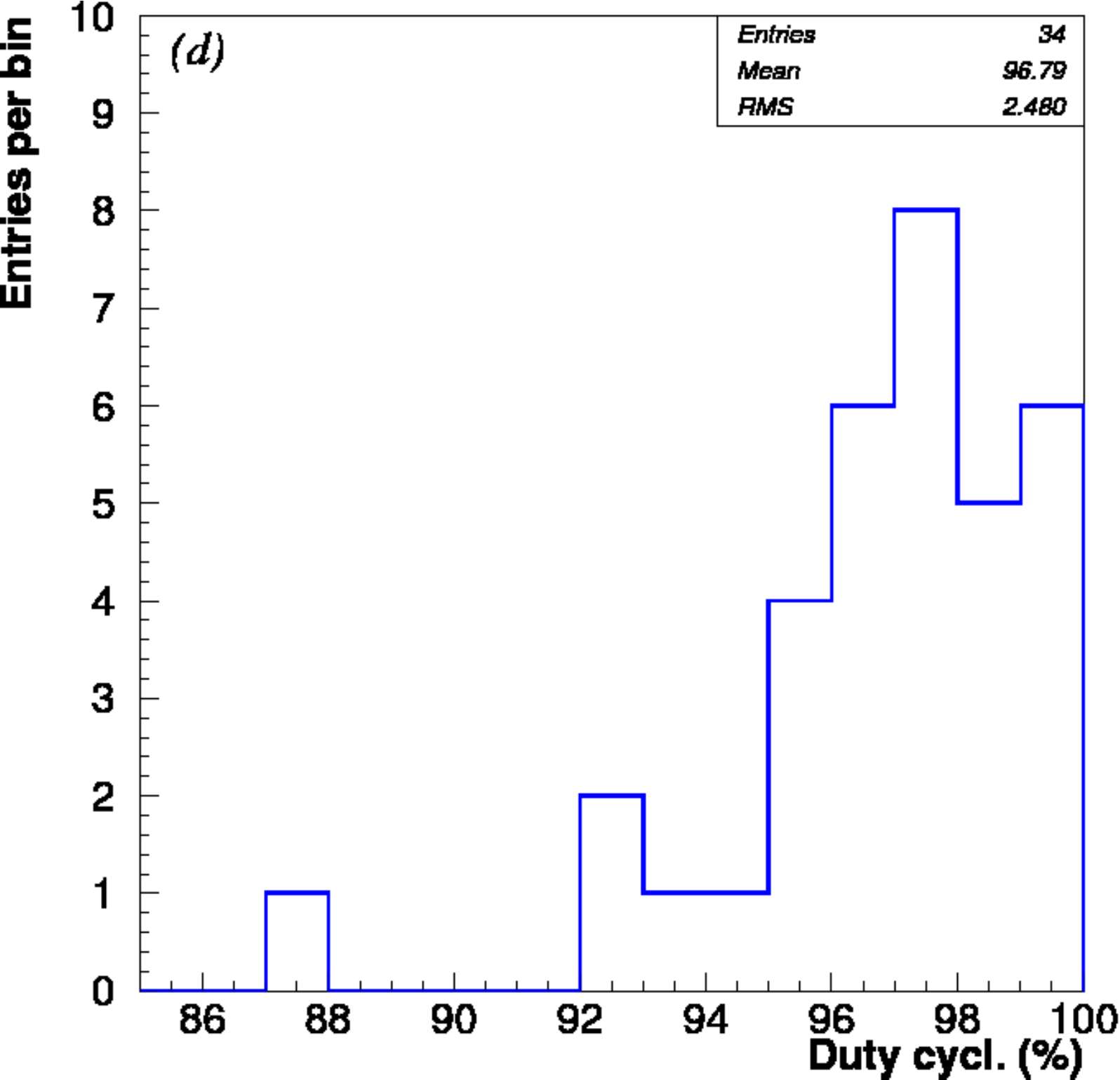}
\end{center}
\vspace{-5mm}
\caption{
Stability of the SK operation: (a) the silent warning rate per 24 hours
as a function of elapsed days from Jan. 1st of 2010,
(b) averaged number of accidental background events per 10 seconds
with one histogram entry for one day, (c) process time to finish the reformat and
reconstruction to issue the silent warnings found, and (d) SN monitor duty
cycle averaged over one month for each entry.
}
\label{fig:stability}
\end{figure}

%\textcolor{blue}{
We use the spallation events found as silent warnings to estimate the false alarm rate 
by assuming constant rate Poisson processes 
rather than generating simulation samples of the backgrounds. 
We combine multiple silent warnings randomly and form a combined cluster 
to estimate a probability of having a golden (normal) warning. 
Using 2,551 silent warnings, we combined two of them for all combinations
of two spallation bursts, to form 
$_{2551}C_2=3,252,525$ patterns, and estimate the probability 
for the combined burst to pass the criteria for a golden (normal) warning to be 0 ($4.3\cdot 10^{-6}$). 
For three combinations with $_{2551}C_3$ patterns, the probability is estimated to be 
$4.1\cdot 10^{-5} (1.9\cdot 10^{-4})$ for a golden (normal) warning. 
One silent warning happens every 10 hours. 
The probability to have two (three) spallation clusters coincident within 20 sec is
%$20/10/3600 =$
$5.6\cdot 10^{-4}~(3.1\cdot 10^{-7} )$. 
%Therefore a probability to have golden (normal) waning with spallation bursts 
%accidentally coincident is 
Therefore the probability to have a golden (normal) warning due to accidentally coincident spallation bursts is
$3.1\cdot 10^{-7} \times 4.1\cdot 10^{-5}=1.3\cdot 10^{-11}
 (5.6\cdot 10^{-4} \times 4.3\cdot 10^{-6} = 2.4\cdot 10^{-9})$.
The false alarm rates are calculated to be once per 
%10 hours / $1.3\cdot 10^{-11} = 
$9.0\cdot 10^7$ years for a golden warning and once per
%10 hours /$2.4\cdot 10^{-9} = 
$4.7\cdot 10^5$ years for a normal warning.
%}

%The processes in the SN burst monitor are kept under observation by a web-based monitor running on a different PC.
%Any problems are displayed on the web monitor immediately after they are found,
%and the SK experiment shift is notified 
%\textcolor{red}{
%by visible and audio alerts.}

The processes in the SN burst monitor are kept under observation 
by a web-based monitor running on a dedicated PC. 
Any problems are displayed on the web monitor immediately 
after they are found, and the SK shift takers are notified 
%\textcolor{blue}{
by visible and audio alerts.
%}

\section{Determination of the SN direction} \label{sec:direction}

The determination of the SN direction is crucial
since the direction information provided by 
neutrinos is useful for astronomers to observe 
the SN explosion process from the onset via electromagnetic waves.
At present, 
SK is the only operating experiment with sufficient detector mass 
to determine the neutrino direction from elastic scattering events
%only SK is able to determine the direction using neutrino events:
%because of the large detector volume, SK can accumulate a sufficient number of
%elastic scattering events, as shown in Table~\ref{table:expected-events},
which preserve the SN direction.
Though the inverse beta decay events also have a correlation with
the SN direction, the elastic scattering events mainly dominate the direction determination power.
A study of the determination of the SN direction using neutrinos was performed by
\cite{sn-pointing}\cite{cc-oxygen2}.
Here, we present a method we have developed.
We will explain the algorithm to determine the SN direction, and then
will show its performance obtained using SK MC.

\subsection{Algorithm}

We determine the SN direction based on a maximum likelihood method.
A likelihood function $L_i$ for $i$-th event is defined as:
\begin{equation}
L_i = \sum_r N_{rk} p_r(E_i, \hat{d}_i; \hat{d}_{\rm SN}),
\end{equation}
where the index $r$ indicates one of the four neutrino interaction channels:
 inverse beta decay ($\bar{\nu}_e p$), 
electron elastic scattering of anti-electron neutrino ($\bar{\nu}_e e$),
other elastic scatterings ($\nu e$)
and the charged-current interactions on oxygen ($\nu ^{16}$O).
%In a fit, we neglect the neutrino interactions to oxygen,
%since those contribution is small ($<0.05$) and has a flat scattering angle distribution.
The index $k$ indicates the energy bin, running from 1 to 5 for
the energy ranges of $7 < E < 10$, $10 < E < 15$, $15 < E < 22$,
$22 < E < 35$ and $35 < E < 50$, respectively,
where $E$ is the measured total electron energy in MeV.
$N_{rk}$ is the number of events of the interaction $r$ in the $k$-th energy bin.
$E_i$ is the $i$-th event total electron energy, which uniquely determines the
index $k$,
$\hat{d}_i$ is the $i$-th event direction and
$\hat{d}_{\rm SN}$ is the SN direction we want to determine.
The $p_r(E_i, \hat{d}_i; \hat{d}_{\rm SN})$ function is a probability density
function (PDF) for interaction $r$ as a function of the energy $E_i$
and an inner-product of $\hat{d}_i \cdot \hat{d}_{\rm SN}=\cos\theta_{\rm SN}$.
The PDF is determined using SK MC.
The number of $\bar{\nu}_e$ elastic scattering events can be
inferred from the number of inverse beta decay events with the relation
$N_{\bar{\nu}_e e,~k} = \sum_m A_{km}N_{\bar{\nu}_e p,~m}$,
where the matrix $A_{km}$ is calculated from a ratio of the total
cross sections of the two interactions.
We determine PDFs for elastic scatterings with the following procedure.
We divide the SN MC sample elastic scattering events generated with the 
Wilson model into one-MeV bins from 7 to 35~MeV. 
For energies greater than 35 MeV, we combined all events into one bin.
Then we fit the $\cos\theta_{\rm SN}$ distribution with the known
SN direction in MC using a model function that is the superposition of
four exponential functions and containing eight parameters for each energy bin.
For a given energy value, we compute the eight parameter values by interpolating
the parameter values of neighboring two energy bins and applying those to
the model function to obtain the PDF value.
A similar procedure is applied to the PDFs for inverse beta decays
and interactions on oxygen to determine the PDF values.
%We determine PDFs of elastic scatterings for the $\nu_e$ and $\nu_x$
%separately using a MC simulation and added them with a weight
%assuming the Wilson model to give the total PDF for elastic scattering
%for each 1~MeV energy bin
%Although the fraction may be different for different models, 
%it would be hard to distinguish the two components for the online direction fit
%with the statistics of the neutrino burst at 10~kpc.
%Therefore we use the PDF determined in that way by regarding the
%weight difference as a second-order effect.
We construct a likelihood 
$\displaystyle{{\cal L}=\exp \left( \sum_{k,r}N_{rk} \right)\prod_i L_i}$,
and maximize it so that:
\begin{equation}
\frac{\partial {\cal L}}{\partial N_{rk}} =
\frac{\partial {\cal L}}{\partial \hat{d}_{\rm SN}} = 0,
\end{equation}
where for $N_{rk}$ we vary $r=\bar{\nu}_e p$, $\nu e$ and $\nu ^{16}$O.
For $r=\nu ^{16}$O, we assume the $\cos\theta_{\rm SN}$ is same for
neutrino and anti-neutrino interactions.
We set $N_{rk}=0$ for $r=\nu ^{16}$O with $k=1,~2,~3$, as the
expected number of charged current interactions on oxygen is negligible
in those energy ranges,
%\textcolor{blue}{
as shown in Fig.~\ref{fig:sk_vis_ene}.%}
The SN direction $\hat{d}_{\rm SN}$ contains two parameters: 
zenith and azimuth angles,
%\textcolor{red}{
that are translated to the direction in the equatorial coordinate system
with the time the burst is found.
%}
Hence, we vary 14 parameters of $N_{rk}$ and $\hat{d}_{\rm SN}$ in total.

When we perform a fit with the likelihood method, we first determine the initial value
of the direction based on a grid search: 
we scan all $\hat{d}_{\rm SN}$ to the $4\pi$ directions 
with a coarse grid step and count the number of
events that satisfy $\cos\theta_{\rm SN} > 0.8$ at each step, and we set the initial value
that gives the largest number of events.

\subsection{Performance}

\begin{figure}[h]
\begin{center}
\includegraphics[width=115mm,clip]{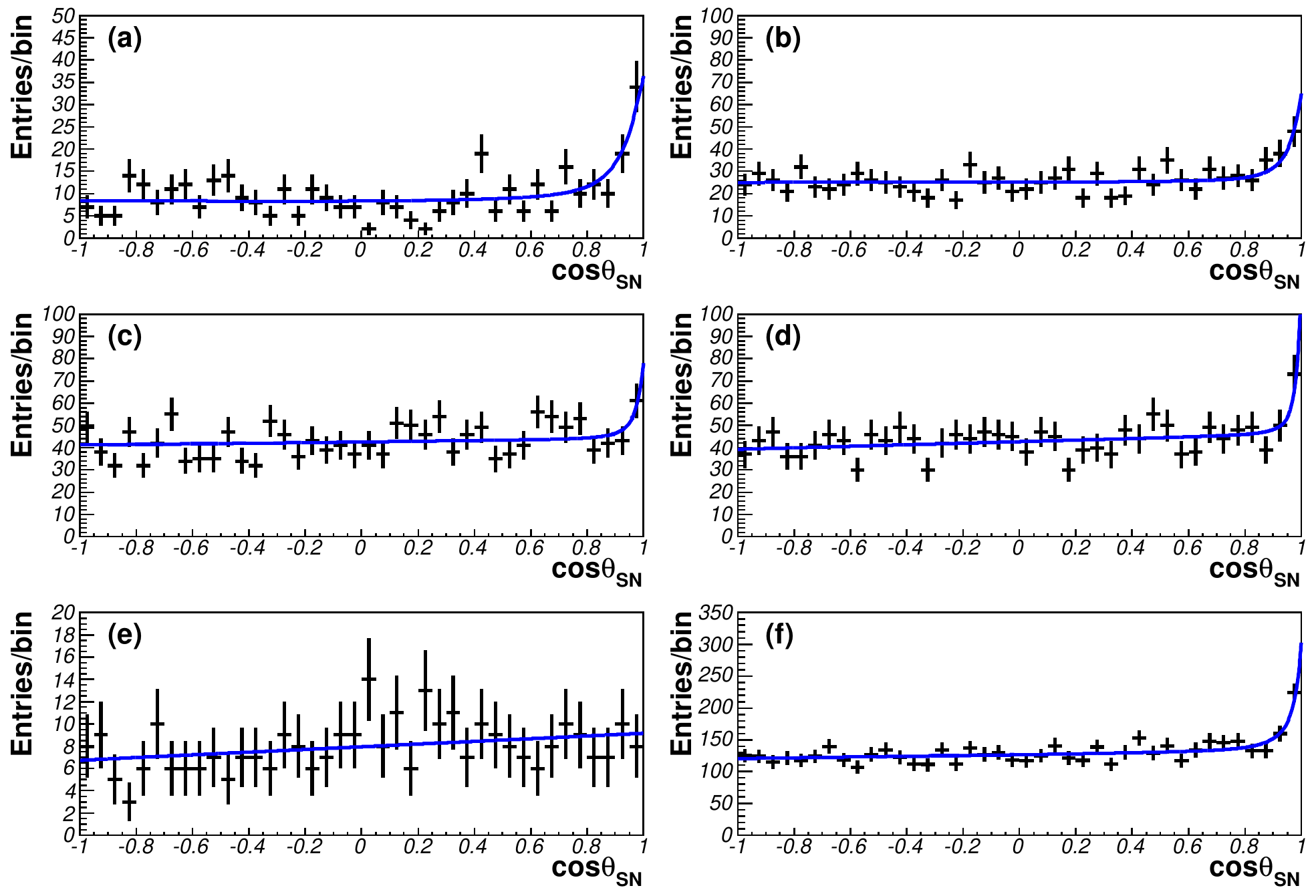}
\end{center}
\vspace{-5mm}
\caption{
Distributions of $\cos\theta_{\rm SN}$ with the Wilson model at 10~kpc
for the five energy bins:
(a)~ $7 < E < 10$~MeV, (b)~$10 < E < 15$~MeV, (c)~$15 < E < 22$~MeV,
(d)~$22 < E < 35$~MeV and (e)~$35 < E < 50$~MeV, where $E$ is the measured total electron 
energy, and (f)~all energies combined.
The superimposed solid lines are the fitted likelihood functions.
}
\label{fig:sn-direction-dist}
\end{figure}

\begin{figure}[h]
\begin{center}
\includegraphics[width=110mm]{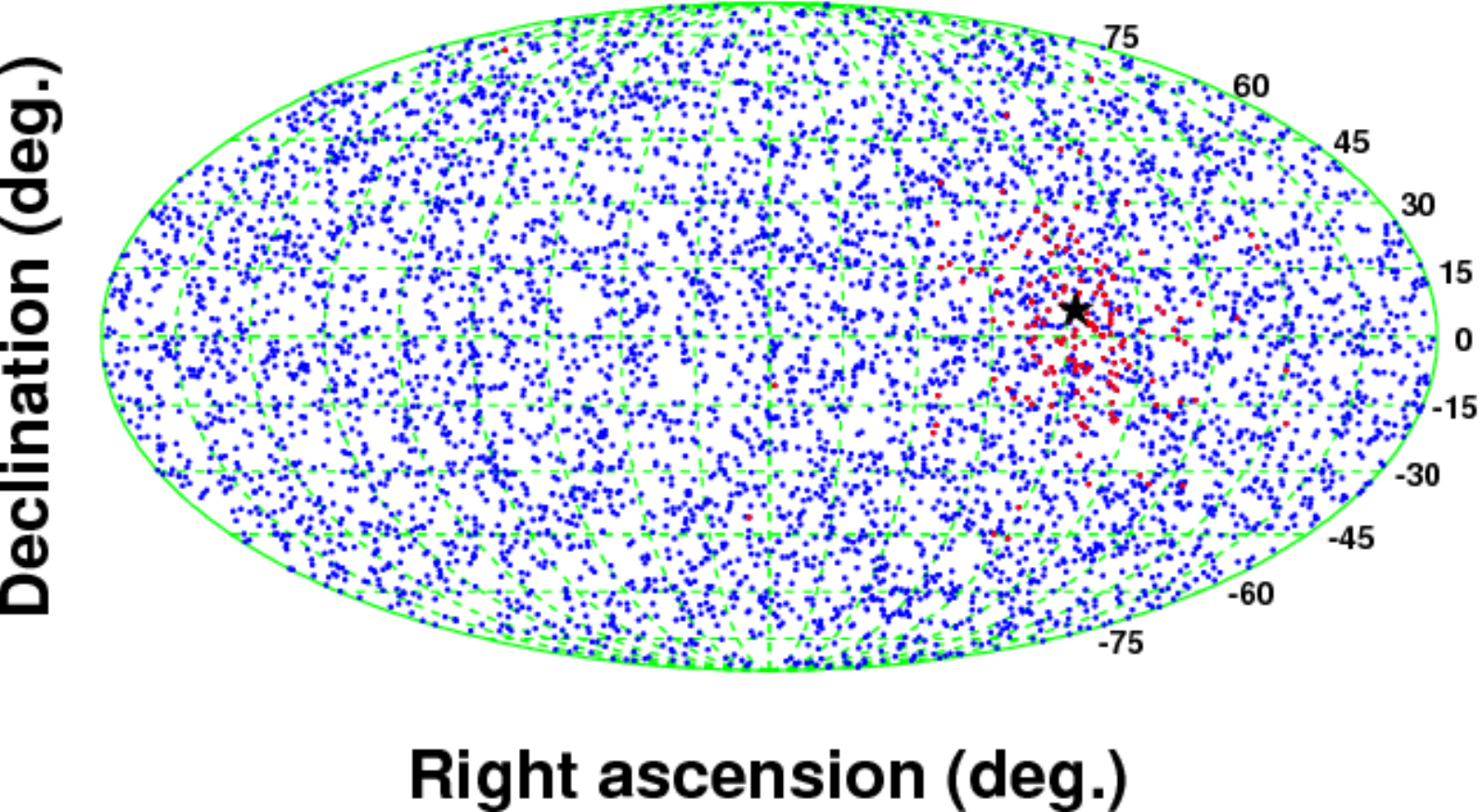}
\end{center}
%\vspace{-5mm}
\caption{
Reconstructed direction on a sky map in the equatorial system
obtained by a MC simulation with the Wilson model at 10~kpc.
Red points are the directions of elastic scattering events,
blue points are event directions of inverse beta decay and 
charged currents on oxygen,
and the star point is the reconstructed SN direction. 
}
\label{fig:skymap}
\end{figure}

Figure \ref{fig:sn-direction-dist} demonstrates $\cos\theta_{\rm SN}$ distributions
of a fit to a MC sample of the Wilson model at 10~kpc
for the five energy bins and combined one
with the superpositions of the fitted likelihood functions.
Figure~\ref{fig:skymap} shows the corresponding direction distribution
on a sky map in the equatorial system.
%\textcolor{blue}{
%The red (blue) points are the reconstructed direction of each event
%of an elastic scattering (an inverse beta decay or a charged current on oxygen), and
%the star mark shows the reconstructed SN direction.
%The elastic scattering events concentrate around the reconstructed
%SN direction, while the inverse beta decays and the charged currents on oxygen
%distribute almost uniformly in all the sky area.} 
The red (blue) points are the reconstructed directions of each elastic scattering event (inverse beta decay or charged current reaction on oxygen), and the star mark shows the reconstructed SN direction. The elastic scattering events concentrate around the reconstructed SN direction, while the distribution of inverse beta decays and charged current events is almost uniform across the entire sky.%}

\begin{figure}[h]
\begin{center}
\includegraphics[width=85mm,clip]{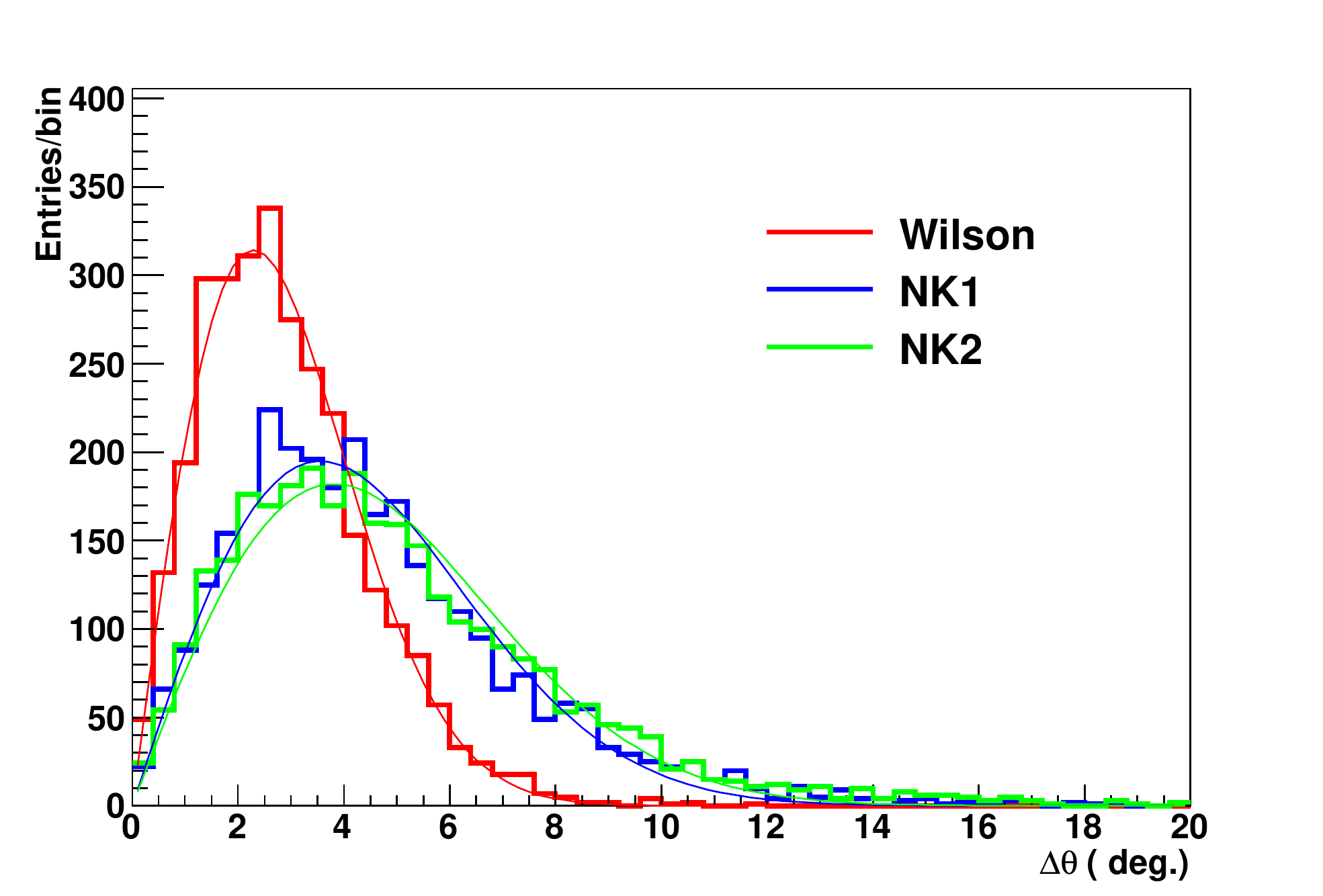}
\end{center}
\vspace{-5mm}
\caption{
Distributions of $\Delta\theta$ for three models at a distance of 10~kpc without
neutrino oscillation.
The superimposed solid lines are the fitted MF functions.
}
\label{fig:dtheta}
\end{figure}

%The reconstructed SN direction is indicated by a red point.
Figure \ref{fig:dtheta} shows $\Delta\theta$ distributions of the three models
(Wilson, NK1 and NK2) for 3,000 MC samples at 10~kpc without neutrino oscillation, 
where $\Delta\theta$ is the angle
between the input SN direction and the fitted direction.
The solid lines are fit results using the von Mises-Fisher (MF) distribution~\cite{MF-dist}:
\begin{equation}
f(\Delta \theta;\kappa) = \frac{\kappa}{2\sinh\kappa}e^{\kappa\cos\Delta \theta}\sin\Delta \theta,
\end{equation}
where $\kappa$ determines the sharpness of the distribution concentration on a sphere.

We estimate the angular resolutions of the SN direction determination using
an ensemble estimation.
In order to cope with any possible combinations of the elastic scatterings and
inverse beta decays, we employ the following method. 
We generate a number of MC samples for various combinations
of fitted yields of the elastic scatterings and inverse beta decays
in the ranges up to 1,500 for the former and 60,000 for the latter.
We divide each range into 15 to obtain a 15$\times$15 matrix.
Each matrix element contains about 3,000 MC samples.
%and divide them into 15$\times$15 elements. Each matrix element
%contains about 3,000 MC samples.
For each element, we determine the
angle $\theta_{\rm en}$ that covers 68.2\%, 90\% and 95\% of the MC samples.
We also provide probabilities to have the true SN position in 2, 5 and 10 degrees with 
respect to the fitted direction.
We generate the matrices for the Wilson and NK1 models with (NH) and without
neutrino oscillations. 
The dependence of the $\theta_{\rm en}$ on the models has about a 10\% variation.
%a variance is about 10%, the min-max ranges about 30%
We employ the largest values of $\theta_{\rm en}$ and the smallest
value of the probabilities for each matrix element conservatively.
When we find a SN neutrino burst, we apply the fit to the burst events to obtain
the SN direction and the yields of the elastic scatterings and inverse beta decays.
With the fitted yields, we identify the matrix element and obtain the angular
resolutions and the probabilities 
%\textcolor{blue}{
that are to be announced to the public.%}
%\textcolor{red}{
For example, we find $\theta_{\rm en} = 3.1\sim3.8^{\circ} (4.3\sim5.9^{\circ})$ at 68.2\% coverage for the Wilson (NK1) model at 10 kpc, where the range covers various neutrino oscillation scenarios in Table~\ref{table:expected-events}.
%}
%For example, we find $\theta_{\rm en}=3.5^{\circ} (5.6^{\circ})$ at 68.2\% coverage
%for the nominal Wilson (NK1) model at 10~kpc without neutrino oscillation.
%We produce the matrix tables for the two models: Wilson and NK1, and 
%employ larger values for each element conservatively 
%to take into account the difference of the energy spectrum.
%In this manner, we are able to give angular resolutions almost independent of
%the SN models.
%and the existence of neutrino oscillations.

The Nakazato model also provides a SN model
with a black hole formation for $M=30$ solar mass~\cite{nakazato-model} .
%\textcolor{blue}{
%The model predicts the neutrino emission that suddenly stops at 842~ms
%after the core bounce.
%In case the SN monitor observes a sudden drop of the neutrino burst,
%that may be a signature of the formation of a black hole.}
The model predicts that neutrino emission suddenly stops 842 ms after the core bounce.  If the SN monitor observes an abrupt cutoff of the supernova neutrino flux, this could be the signature of the birth of a black hole.%}
Based on this model, we generate MC samples and apply the fit.
We obtain an angular resolution of 2.3 degrees at 10 kpc with 68.2\% confidence
level.
Therefore this may help the identification of a position of a 
disappeared massive star as proposed by ~\cite{survey-nothing}.

In order to understand the behavior of the estimated angular resolution,
we make use of the curvature of the likelihood at its maximal position:
we define a value $\sigma$ as:
\begin{equation}
\sigma = \sqrt{-\frac{1}{\displaystyle{\frac{\partial^2 \ln {\cal L}}{\partial \theta_{\rm SN}^2}}}},
\end{equation}
where the second derivative is the curvature calculated at the point on the sphere at which
the likelihood function ${\cal L}$ becomes maximal.
We calculate $\sigma$ values along four planes that include the maximal point and have different
azimuth angles with respect to the point,
and employ the maximum among the four $\sigma$ values.
Then we obtain the resolution $\theta_{\sigma}(q)$ that covers an area with a fraction of $q = 1-p$
of the MF distribution with a $p$ value:
\begin{equation}
\theta_{\sigma}(q)=\arccos\left[\frac{1}{\kappa}
\ln\left(1-q+q e^{-2\kappa}\right)+1\right], \label{eq:mf-func-integ}
\end{equation}
and $\kappa = 1/\sigma^2$.

\begin{figure}[h]
\begin{center}
\includegraphics[width=65mm,clip]{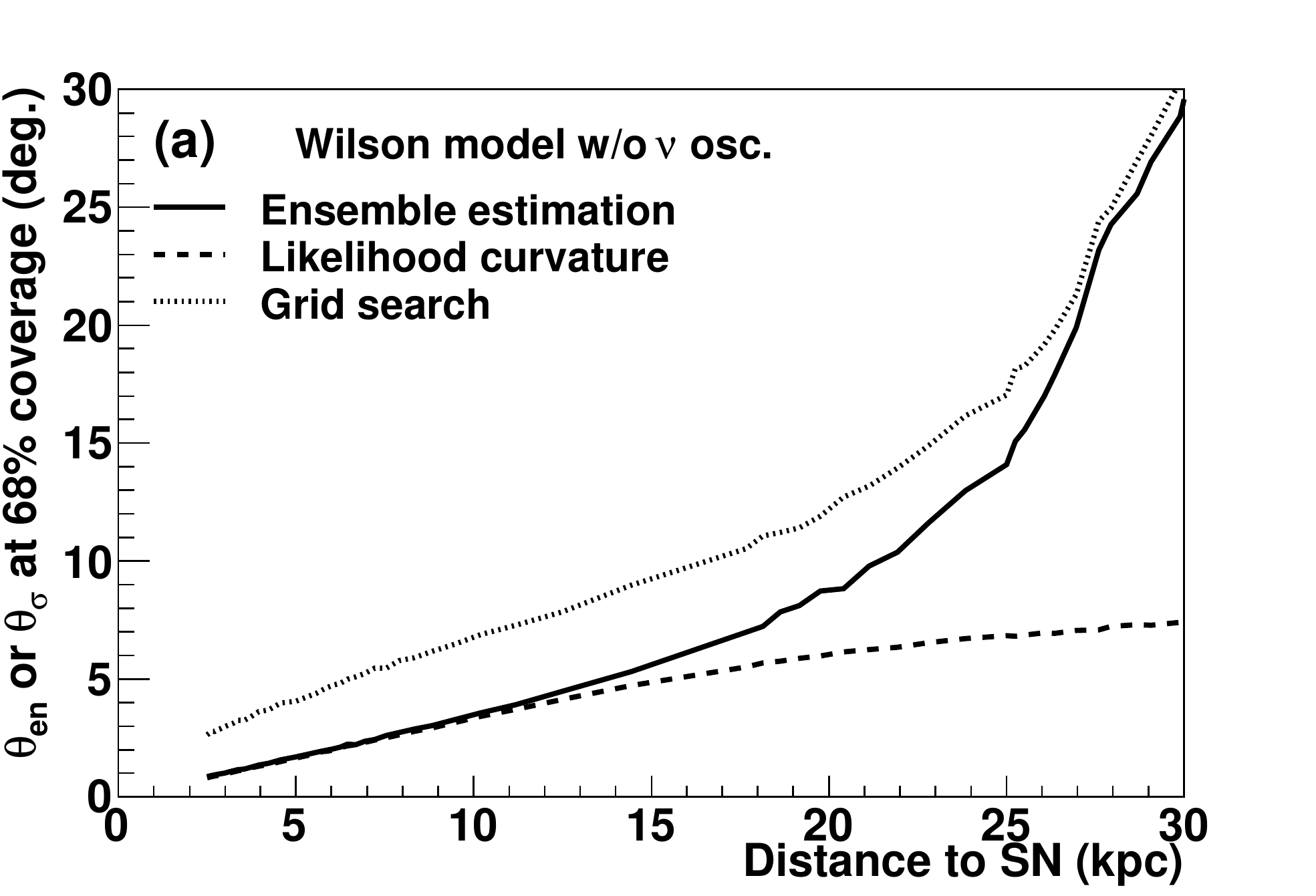}
\includegraphics[width=65mm,clip]{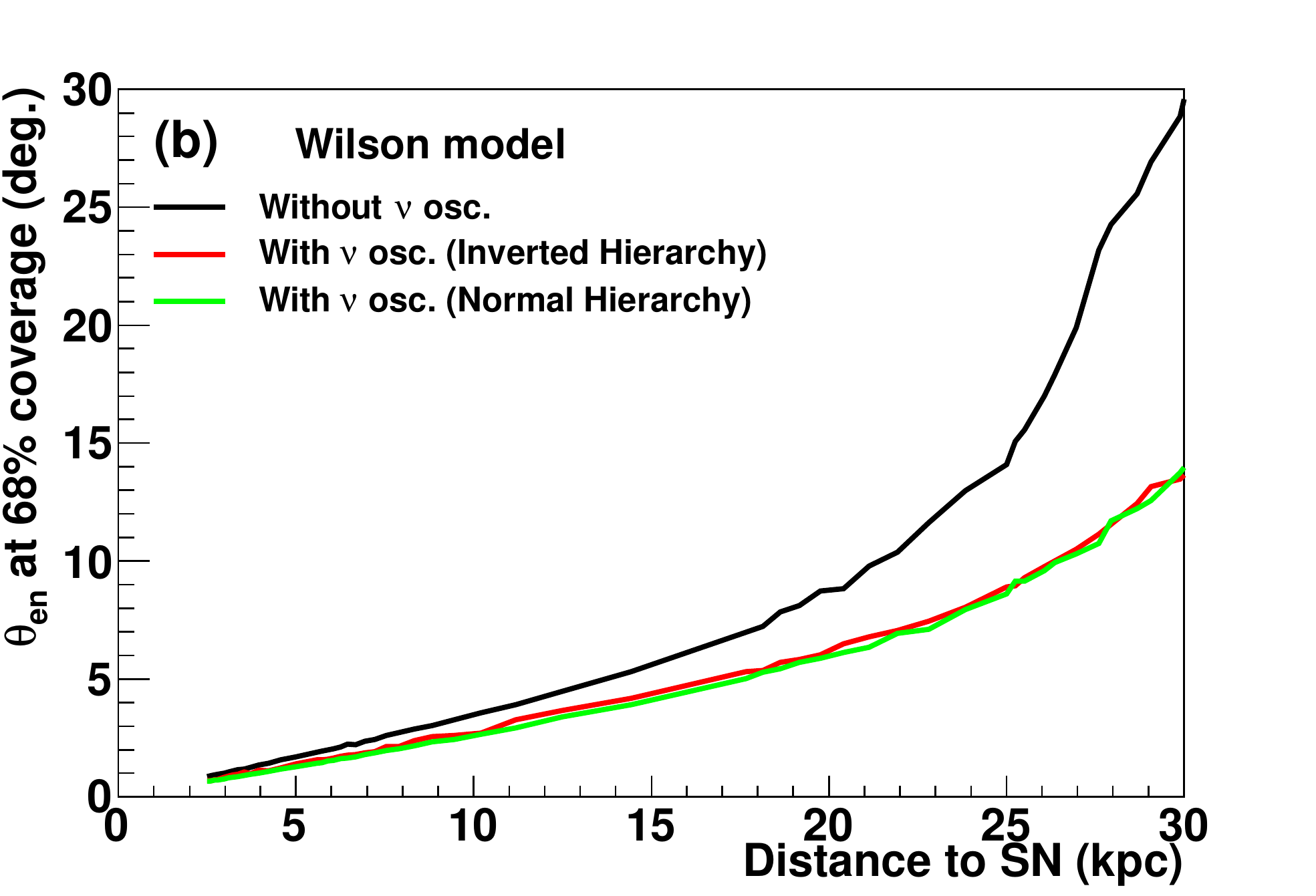}
\includegraphics[width=65mm,clip]{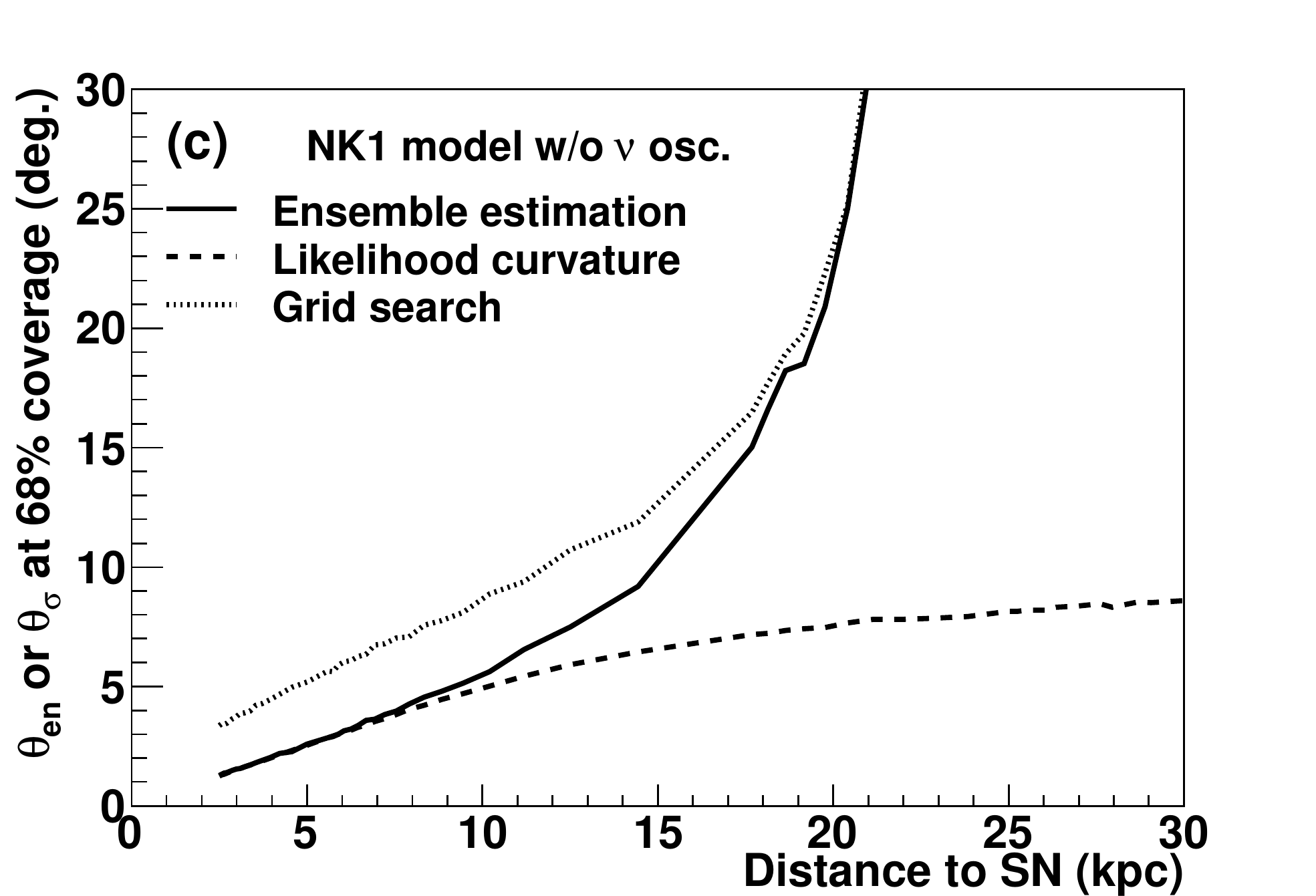}
\includegraphics[width=65mm,clip]{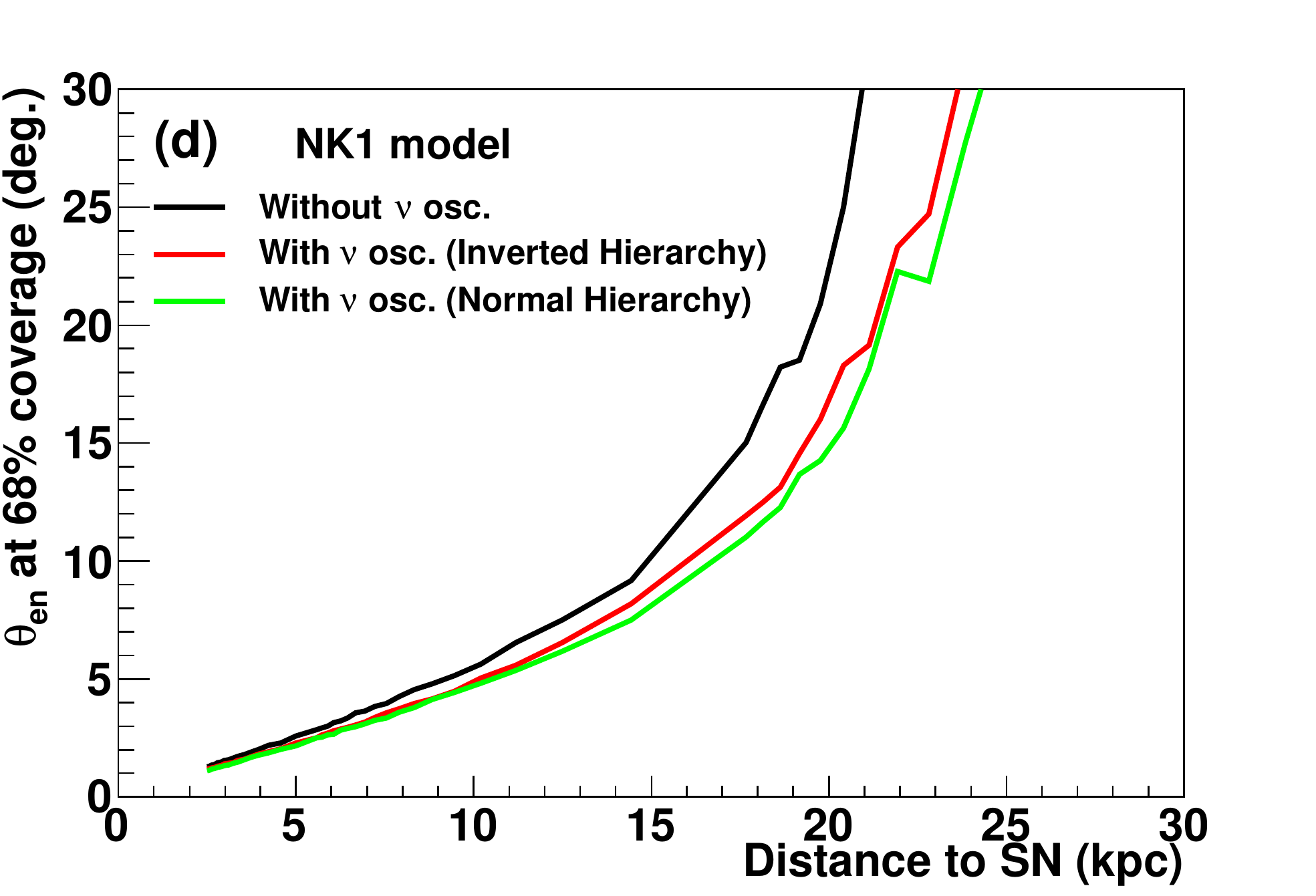}
\end{center}
\vspace{-5mm}
\caption{
$\Delta\theta$ that covers 68\% of SNe (the area) as a function of distance.
In (a) and (c), the solid, dashed and dotted lines correspond to 
$\theta_{\rm en}$ obtained by the ensemble estimation,
$\theta_{\sigma}(0.682)$ obtained by the likelihood curvature 
with Eq.(\ref{eq:mf-func-integ}),
and that obtained by the grid search, respectively,
for the Wilson (a) and NK1 (c) models.
Figure (b) and (d) show $\theta_{\rm en}$ as a function of distance for
the three neutrino oscillation hypotheses for Wilson (b) and NK1 (d) models.  
}
\label{fig:dtheta_dist}
\end{figure}

In Fig.~\ref{fig:dtheta_dist} (a) and (c), we show the obtained angular resolution at 
the 68.2\% coverage
for the ensemble estimation of the fit, the likelihood curvature method
and the ensemble estimation using a grid search as a function of the distance
for the Wilson and NK1 models.
We find that the $\Delta\theta$ distributions are well modeled by the MF function
and the likelihood curvature estimation $\theta_{\sigma}(q)~(q=0.682)$
is consistent with that of the ensemble estimation
for the statistics larger than those of SNe at 10~kpc (7~kpc) for the Wilson (NK1) model.
However, $\theta_{\sigma}(q)$ is found to be an underestimate for the smaller statistics.
The degradation of the angular resolution for the distant SNe is likely
due to failure in giving a proper initial value of the SN direction by the 
grid search, indicated by the fact that
the $\theta_{\rm en}$ value approaches 
that of the grid search for larger distances (smaller statistics).
The small statistics produce a large fluctuation that sometimes makes a fake peak
on the grid direction search.
That makes the initial value a wrong direction and the likelihood fit finds 
a local minimum around the direction. 
%In this case, the fit finds a local minimum around the given initial value.
%The probablity to have a fake minimum in the fit increases as the statistics decrease.
Figure~\ref{fig:dtheta_dist} (b) and (d) show the ensemble estimation
of the angular resolution as a function of the SN distance for three 
neutrino oscillation hypotheses. 
The angular resolutions with the two neutrino oscillation hypotheses are
smaller than those without neutrino oscillation.
This is due to an increase in elastic scattering events
as shown in Table~\ref{table:expected-events}.

%\textcolor{blue}{
We estimate the precision of the angular uncertainty using Eq.~(\ref{eq:mf-func-integ})
under the assumptions of the SN models used; 
the precision of $\Delta \theta_{\sigma}$ is determined using $\Delta \theta_{\sigma}
 = d\theta_{\sigma} / dq \cdot \Delta q = d\theta_{\sigma} / dq \cdot \sqrt{q/ N}$,
where $N=3,000$ is the number of samples in the ensemble. 
We find $\Delta \theta_{\sigma} = 0.09$ degrees with $q=0.682$
for the NK1 model with the NH neutrino oscillation at a distance of 10 kpc, 
which is much smaller than the angular resolution variation of various 
neutrino oscillation scenarios.
%}

%It is quite important to give a radius of an error circle along with the SN direction.
%The $\theta_{\rm en}$ value shown in Fig.~\ref{fig:dtheta_dist} is one candidate.
%However, the value has a dependence not only on the SN models but also on
%the neutrino oscillation parameters and hierarchy hypotheses.
%In order to copy with any SNe, we have decided to provide a two dimensional table that contains
%$\theta_{\rm en}$ values as a function of numbers of elastic scattering events and
%inverse beta decays (plus Oxygen events) obtained by a fit.
%Even this, we may still have a model dependence because of the differences of energy spectra.
%As a benchmark, we make the tables using the Wilson and NK1 models and choose larger values
%of each element as a conservative way.

%For larger statistics, it is not easy to estimate $\theta_{\rm en}$ since this method
%requires a huge computational power.
%Instead, we employ $\theta_{\sigma}(q)$ value for SNe 
%with the fitted number of elastic scattering events
%greater than 1000 corresponding to about 3~kpc for the NK1 model.

\section{Summary}

We describe a real-time monitor of a SN neutrino burst at SK. 
The monitor is able to provide a fast warning to the world within one hour.
The system is operating on a dedicated computer independent 
of the offline processes.
%with no dead time.
The SN neutrino burst selection criteria are determined so that
fake event bursts mainly caused by spallation events are rejected.
Using MC simulations, we find that the system has 100\% detection efficiency up to the LMA
for the three SN models with the golden warning criteria.
The expected total number of neutrino events with the selection criteria
is about 5,200 (2,200) for the Wilson (NK1) model at 10~kpc without neutrino oscillation.
Neutrino oscillations increase the number of inverse beta decay events
and enhance the detection efficiency for SNe at the SMC 
for both normal and inverted hierarchy hypotheses.

The SN direction pointing is of importance as it provides a chance for astronomers
to observe the SN explosion from its onset with electromagnetic waves.
SK is the only detector that enables us to reconstruct the SN direction using only neutrinos
among the existing neutrino observatories.
We have developed an algorithm to identify the SN direction and its error using
a maximum likelihood method.
%\textcolor{red}{
The pointing accuracy estimated by the ensemble study is found to be 
$3.1\sim3.8^{\circ} (4.3\sim5.9^{\circ})$ at 68.2\% coverage for the Wilson (NK1) 
model at 10 kpc, where the range covers various neutrino oscillation scenarios.
%}
%The pointing accuracy estimated by the ensemble study is found to be
%$3.5^{\circ} (5.6^{\circ})$ at 68\% coverage 
%for the Wilson (NK1) model
%at a distance of 10~kpc, where the range covers various neutrino oscillation
%scenarios.

\section*{Acknowledgements}

We gratefully acknowledge the cooperation of the Kamioka Mining and Smelting Company.
The Super-Kamiokande experiment has been built and operated from funding by the
Japanese Ministry of Education, Culture, Sports, Science and Technology, the U.S.
Department of Energy, and the U.S. National Science Foundation. Some of us have been
supported by funds from the Research Foundation of Korea (BK21 and KNRC), the Korean
Ministry of Science and Technology, the National Research Foundation of Korea (NRF-
20110024009), the European Union (H2020 RISE-GA641540-SKPLUS), the Japan Society for
the Promotion of Science, the National Natural Science Foundation of China under Grants
No. 11235006, the National Science and Engineering Research Council (NSERC) of Canada,
and the Scinet and Westgrid consortia of Compute Canada.
This work was partly supported by the Grant-in-Aid for Scientific
Research on Innovative Areas [JSPS No.26104006].



%% References


\bibliographystyle{elsarticle-num}
\bibliography{<your-bib-database>}



\end{document}